\documentclass[aps,pra,floatfix,superscriptaddress,notitlepage,twocolumn]{revtex4-2}

\usepackage[T2A]{fontenc}
\usepackage[utf8]{inputenc}
\usepackage[english]{babel}
\usepackage[colorlinks=true, allcolors=blue]{hyperref}
\usepackage{bm}
\usepackage{graphicx}
\usepackage{amsmath}
\usepackage{array}
\usepackage{multirow}

\begin{document}

\title{Thermal one-loop self-energy correction for hydrogen-like systems: relativistic approach}

\author{M.~A.~Reiter} 
\email[E-mail:]{ mikh.reiter@gmail.com}
\affiliation{Department of Physics, St. Petersburg State University, Universitetskaya nab. 7/9, 199034 St. Petersburg, Russia}

\author{D.~A.~Solovyev}
\affiliation{Department of Physics, St. Petersburg State University, Universitetskaya nab. 7/9, 199034 St. Petersburg, Russia}
\affiliation{Petersburg Nuclear Physics Institute named by B.P. Konstantinov of NRC ``Kurchatov Institute'', Orlova roscha 1, 188300 Gatchina, Leningrad region, Russia}

\author{A.~A.~Bobylev}
\affiliation{Department of Physics, St. Petersburg State University, Universitetskaya nab. 7/9, 199034 St. Petersburg, Russia}
\affiliation{Petersburg Nuclear Physics Institute named by B.P. Konstantinov of NRC ``Kurchatov Institute'', Orlova roscha 1, 188300 Gatchina, Leningrad region, Russia}
   
\author{D.~A.~Glazov}
\affiliation{School of Physics and Engineering, ITMO University, Kronverkskiy pr. 49, 197101 St. Petersburg, Russia}
\affiliation{Petersburg Nuclear Physics Institute named by B.P. Konstantinov of NRC ``Kurchatov Institute'', Orlova roscha 1, 188300 Gatchina, Leningrad region, Russia}

\author{T.~A.~Zalialiutdinov}
\affiliation{Department of Physics, St. Petersburg State University, Universitetskaya nab. 7/9, 199034 St. Petersburg, Russia}
\affiliation{Petersburg Nuclear Physics Institute named by B.P. Konstantinov of NRC ``Kurchatov Institute'', Orlova roscha 1, 188300 Gatchina, Leningrad region, Russia}

\begin{abstract}

Within a fully relativistic framework, the one-loop self-energy correction for a bound electron is derived and extended to incorporate the effects of external thermal radiation. In a series of previous works, it was shown that in quantum electrodynamics at finite temperature (QED), the description of effects caused by blackbody radiation can be reduced to using the thermal part of the photon propagator. As a consequence of the non-relativistic approximation in the calculation of the thermal one-loop self-energy correction, well-known quantum-mechanical (QM) phenomena emerge at successive orders: the Stark effect arises at leading order in $\alpha Z$, the Zeeman effect appears in the next-to-leading non-relativistic correction, accompanied by diamagnetic contributions and their relativistic refinements, among other perturbative corrections. The fully relativistic approach used in this work for calculating the SE contribution allows for accurate calculations of the thermal shift of atomic levels, in which all these effects are automatically taken into account. The hydrogen atom serves as the basis for testing a fully relativistic approach to such calculations. Additionally, an analysis is presented of the behavior of the thermal shift caused by the thermal one-loop correction to the self-energy of a bound electron for hydrogen-like ions with an arbitrary nuclear charge $Z$. The significance of these calculations lies in their relevance to contemporary high-precision experiments, where thermal radiation constitutes one of the major contributions to the overall uncertainty budget.


\end{abstract}

\maketitle

\section{Introduction}

Given the modern development of spectroscopic experiments and the significantly increased accuracy in determining transition frequency standards over recent decades, increasingly smaller effects in terms of their absolute value are beginning to play a key role. To achieve such progress, the evolution of quantum theory and the corresponding methods and approaches was required. The main tool in the study of the spectroscopic properties of atomic systems is the theory of quantum electrodynamics (QED) \cite{Greiner,AB}, which allows for an unprecedented comparison of theoretical and experimental results. A clear example for such analysis can be the hydrogen atom, the accuracy of measurement of the transition frequency in which reaches an order of $10^{-15}$ relative value, see, e.g., \cite{CODATA-2021}. Attention can also be drawn to the use of various hydrogen-like ions and other one-electron systems, with which such physical quantities as the g-factor of a bound electron and consequently the electron mass \cite{Sturm2014}, the proton-to-electron mass ratio, etc., \cite{Karr_2025} have been determined with a percentage per trillion level of accuracy.

Covering a wide range of tasks, the comparative analysis of theory with experimental measurement results typically reduces to calculating QED corrections of the next order of smallness (within various perturbation theories, different smallness parameters may be used). At present, it can be stated that for light atomic systems, theoretical research already deals with corrections of the order of $\alpha^7$ ($\alpha$ is the fine structure constant), see, for example \cite{Korobov2014,PPY-2017,CODATA-2021}. Aiming for precise determination of fundamental physical constants, verification of fundamental interactions, or investigation of physical symmetries violations, this kind of analysis occasionally reveals differences that are then used to construct theoretical hypotheses beyond the framework of the Standard Model \cite{PPY-2017,PhysRevA.98.022501,Kozlov-RevModPhys,Blaum_2021}.

However, the primary standard for achieving the highest accuracy in spectroscopic measurements is atomic clocks, as well as various candidates for this role, see \cite{Kozlov-RevModPhys} and the references therein. Atomic clocks, which allow for highly precise measurements, have become a sensitive tool for detecting changes in the gravitational field \cite{Bothwell_2022}, testing global symmetry \cite{Sanner2019}, or imposing constraints on dark matter \cite{DM-atcl}. Experimental precision at the level of $10^{-18}$ and beyond has necessitated theoretical studies of effects that cause even the slightest shifts in atomic energy levels, and also  disrupt the stability of the operational transition in atomic clocks.

It is precisely these effects that reflect the influence of the thermal environment on atomic characteristics. Blackbody radiation (BBR) was the first such influence to be observed experimentally~\cite{GC}, with estimates for the resulting frequency shifts and transition rates provided therein. A more rigorous quantum-electrodynamical treatment followed in~\cite{farley}. The theory and approaches presented in \cite{farley} for calculating the thermal Stark effect and the corresponding transition probabilities induced by BBR are fundamental even today. Such effects have drawn considerable attention in recent decades, effectively becoming a distinct area of theoretical research. Forming one of the dominant contributions to the overall experimental error in determining physical quantities, efforts have been directed towards both the most precise calculations (see, e.g., the works by M. Safronova and co-authors \cite{safronova_b_al_in,AtCl-Sr,Saf-Nat} as well as more resent \cite{Beloy_2014,PhysRevA.110.043108,Beloy_2025}), and the search for ways to mitigate the influence of BBR \cite{Yudin,Zuhrianda,PhysRevA.100.023417}.

Despite various possibilities for suppressing thermal shifts of atomic energy levels, the main theoretical task remains their accurate calculation. Even to offset the dominant contribution requires meticulous control of temperature fluctuations and, as a consequence, determining the resulting uncertainties. As noted above, the primary approach for such calculations is based on quantum mechanical theory \cite{farley}. This approach, extended to many-body perturbation theory, has been successfully applied to estimate the dc-Stark shift and its dynamic corrections induced by blackbody radiation \cite{Saf-Nat,Sahoo-AtCl}. However, around the same time as the development of QM theory of BBR influence on atomic characteristics, the theory of quantum electrodynamics at finite temperature (TQED) was also evolving \cite{Dol,Don,DHR}. This theory encompasses a much broader range of problems related to the thermal environment, and allows for a precise description of phenomena even outside of a perturbative regime with respect to temperature \cite{Soto,Soto2}.

In this work, focusing on atomic bound states, we consider temperature ranges relevant to laboratory conditions—specifically, those well below the ionization threshold of the atomic electron. Previously, it was shown that to obtain the well-known quantum mechanical (QM) results for the ac-Stark shift and the depopulation rates of atomic levels, it is sufficient to take into account the thermal one-loop self-energy (SE) correction for the bound electron \cite{PRA_2015,S_2020}. Then the dominant contribution (Stark shift) arises within the nonrelativistic limit for the operators written in the dipole approximation (longwave approximation). Then the dominant contribution (dynamic Stark shift) arises in the non-relativistic limit for the operator written in the dipole approximation (long-wavelength approximation). 

Demonstrating excellent agreement with the QM approach, the TQED result for the SE correction, nonetheless, is not limited to just the Stark shift. A comprehensive analysis of this approach was presented very recently in work \cite{Lopez_2025}. Specifically, it was found that in addition to the dipole dynamic polarizability of the atom, this correction also contains higher multipole polarizabilities, including magnetic, quadrupole, etc. (similar to \cite{Porsev}), as well as relativistic corrections arising from the wave functions and the expansion of the self-energy operator. Thus, a fully relativistic approach to calculating the thermal one-loop self-energy correction for a bound electron appears more advantageous for obtaining an accurate value of the atomic energy level shift. Additionally, it's worth noting that the TQED approach readily allows for the consideration of even more subtle effects, whether they relate to hyperpolarizabilities or corrections to physical quantities such as the g-factor or hyperfine structure in various atomic systems, see, e.g., \cite{china-BBR,ZGSg_2022,ZGS_2022,ZKS_2023,ZS_2023}.

In this study, we provide a comprehensive analysis of the thermal one-loop self-energy correction of a bound electron within a fully relativistic approach. Discarding the imaginary part (corresponding to the BBR-induced level width \cite{PRA_2015}), we focus on precise calculations of the real part of such a correction for the hydrogen atom. Using the partial wave decomposition method, we analyze the behavior of this thermal correction as the nuclear charge $Z$ changes. Calculations are presented for various states of the hydrogen atom (up to a principal quantum number of $n=12$ and an orbital quantum number of $l\leq 3$) at room temperature, and for various temperatures for states that more closely correspond to the most accurately measured transition frequencies \cite{CODATA-2021}. Throughout this article the relativistic units ($\hbar = c = m = 1$, where $m$ is the electron mass) are used.

\section{Theory}

\subsection{QED description of thermal shifts and line broadening}

\begin{figure}[h!]
\centering
\includegraphics[width=0.2\textwidth]{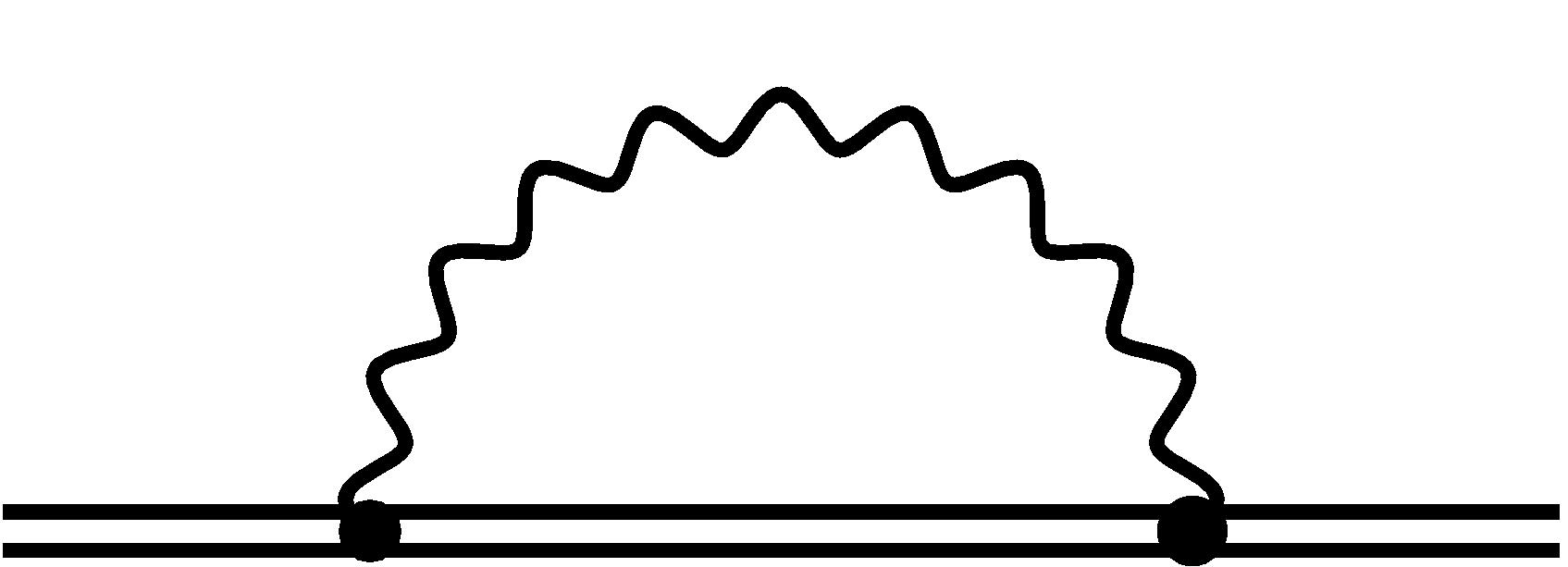}
\caption{One-loop self-energy Feynman diagram for bound electron. The double line indicates the electron propagator in the external field of the nucleus, the wavy line denotes the photon propagator.}
\label{fig:se}
\end{figure}

Referring to the QED description of the atomic level energy shift caused by BBR, we rely on the main conclusions drawn in \cite{Lopez_2025}. According to this theory, the most significant thermal contributions can be obtained by considering the one-loop self-energy correction of the bound electron, in which the ordinary Feynman photon propagator is replaced by the corresponding thermal part~\cite{Dol,Don,DHR}. Specifically, it was shown in \cite{PRA_2015} that the real part of such correction provides the dominant contribution to the atomic level energy shift (Stark shift), while the imaginary part corresponds to the line broadening caused by blackbody radiation. The Feynman diagram showing the thermal one-loop SE correction is illustrated in Fig.~\ref{fig:se}.

Contribution to the energy of a bound level can be obtained within the $S$-matrix formalism. The corresponding matrix element can be written as
\begin{eqnarray}
\langle f | \hat{S}^{(2)}| i \rangle = (-\mathrm{i} e)^2 \int d^4 x_1 d^4 x_2 {\overline{\psi}}_f(x_1) \gamma^\mu S(x_1, x_2)
\\
\nonumber
\times D_{\mu \nu}(x_1, x_2) \gamma^\nu \psi_i(x_2),
\label{eq:s_2}
\end{eqnarray}
where the integration over spacetime variables, $x_1=(t_1, \mathbf{r}_1)$ and $x_2=(t_2, \mathbf{r}_2)$, is implied ($t$ denotes the time variable and $\mathbf{r}$ is the spatial vector). The initial and final states are given by the indexes $i$ and $f$, respectively, $\gamma^\mu$ are the Dirac matrices, $\psi_i(x)$ is the one-electron Dirac wave function, and $\overline{\psi}$ is its Dirac adjoint. For an arbitrary state $|a\rangle$ it can be written as $\psi_a(x) = \psi_a(\mathbf{r}) e^{-{\mathrm{i}} \varepsilon_a t}$ in case of the time-independent Hamiltonian, $\varepsilon_a$ is the corresponding Dirac energy. 

In the following, we consider hydrogen-like ions for which the potential is spherically symmetric. Then, the Dirac wave functions can be represented in the form:
\begin{eqnarray}
\psi_a(\mathbf{r}) = {\begin{pmatrix}g_a(r) \chi_{\kappa_a m_a} (\hat{{\mathbf{r}}}) \\ \mathrm{i} f_a(r) \chi_{-\kappa_a m_a} (\hat{{\mathbf{r}}}) \end{pmatrix}},
\label{wf_coord}
\end{eqnarray}
where $g_a(r)$ and $f_a(r)$ are the large and small radial components, $r=|\mathbf{r}|$, ${\hat{{\mathbf{r}}}}=\mathbf{r}/r$, and $\chi_{\kappa_a m_a} ({\hat{{\mathbf{r}}}})$ is the spin-angular spinor. The subscripts are the Dirac angular quantum number $\kappa_a$ and the projection of the total angular momentum $m_a$. The radial wave functions of the bound-electron state can be characterized by, e.g., $\kappa_a$ and the radial quantum number $n_{r_a}$: $g_a(r)=g_{\kappa_a, n_{r_a}}$, $f_a(r)=f_{\kappa_a, n_{r_a}}$ \cite{AB}.

In Eq.~(\ref{eq:s_2}), $S(x_1, x_2)$ denotes the electron propagator, which admits an eigenmode expansion in terms of the bound-state spectrum, as detailed in, e.g., \cite{Greiner}:
\begin{eqnarray}
\label{3}
S(x_1, x_2) = \frac{\mathrm{i}}{2 \pi}\int\limits_{-\infty}^{+\infty} d \Omega \; e^{-{\mathrm{i}} \Omega(t_1-t_2)} \sum_n \frac{\psi_n(\mathbf{r_1}) {\overline{\psi}}_n (\mathbf{r_2})}{\Omega - \overline{\varepsilon}_n},
\end{eqnarray}
where $\overline{\varepsilon}_n$ is a shorthand notation for $\varepsilon_n(1 - \mathrm{i}\epsilon)$ (with $\epsilon$ being an infinitesimally small positive quantity, regulating the pole bypass rule). The summation in the equation~(\ref{3}) is performed over the entire Dirac spectrum, including the negative continuum.

In turn, the photon propagator $D_{\mu\nu}(x_1, x_2)$ can be used in the form \cite{PRA_2015,S_2020}:
\begin{eqnarray}
\label{phprop}
D_{\mu\nu}(x_1, x_2) = \frac{g_{\mu\nu}}{2 \pi \mathrm{i} r_{12}} \int_{-\infty}^{+\infty} d\omega e^{{\mathrm{i}} |\omega| r_{12} - {\mathrm{i}} \omega (t_1-t_2)} 
\\
\nonumber
-\frac{g_{\mu\nu}}{\pi r_{12}} \int_{-\infty}^{+\infty} d\omega n_\beta(|\omega|) \sin |\omega| r_{12} e^{-{\mathrm{i}} \omega (t_1-t_2)}.
\end{eqnarray}
Here $r_{12} = |\mathbf{r}_1-\mathbf{r}_2|$, $g_{\mu\nu}$ is the metric tensor, and $n_\beta(\omega)$ is given by the Planck distribution function
\begin{eqnarray}
\label{nb}
n_\beta(\omega) = \frac{1}{e^{\omega \slash k_{\rm B} T} - 1}.
\end{eqnarray}
In Eq.~(\ref{nb}) the Boltzmann’s constant is denoted by $k_{\rm B}$ and $T$ represents the temperature of the blackbody radiation in Kelvin. The inverse temperature in units of the Boltzmann constant is usually expressed as $\beta=1/k_{\rm B} T$.

The substitution of expression~(\ref{phprop}) into Eq.~(\ref{eq:s_2}) leads to two components. The first (which corresponds to the photon propagator at zero temperature) determines the SE contribution to the Lamb shift and the radiative level width of the lowest-order (natural). The second component corresponds to the photon propagator at a finite temperature. 

Omitting intermediate calculations for brevity (see, for example, ~\cite{PRA_2015}), the following expression for the thermal SE correction can be obtained:
\begin{eqnarray}
\label{SE_general}
\Delta \varepsilon_a^\beta = \frac{e^2}{\pi} \sum_n \int_{-\infty}^{\infty} d \omega \,n_\beta(|\omega|) \frac{ \langle a n | I(|\omega|) | n a \rangle}{\overline{\varepsilon}_n - \varepsilon_a + \omega},
\end{eqnarray} 
where the operator $I(\omega) \equiv I(\omega, \mathbf{r}_1, \mathbf{r}_2)$ is
\begin{equation}
I(\omega, \mathbf{r}_1, \mathbf{r}_2)= \frac{1 - {\pmb{\alpha}}_1 \cdot {\pmb{\alpha}}_2 }{r_{12}} \sin \omega r_{12}.
\label{op:I}
\end{equation}
Here ${\pmb{\alpha}}_i$ ($i=1,2$) are the Dirac alpha matrices acting in the $i$-th coordinate space. Taking into account the imaginary additive in the energy denominator, using the Sokhotski–Plemelj theorem, it is easy to show that the Stark shift arises from the real part, $\Re \Delta \varepsilon_a^\beta$, and the line broadening caused by BBR from the imaginary part, $\Im \Delta \varepsilon_a^\beta$, of  Eq.~(\ref{SE_general}), see \cite{PRA_2015,S_2020,Lopez_2025}.

\subsection{Partial-wave expansion}

To perform the rigorous relativistic calculations (in contrast to the non-relativistic limit and its extensions as in \cite{Lopez_2025}), we use the partial wave decomposition, see~\cite{abr}, for the operator $I(\omega)$:
\begin{eqnarray}
\label{PE}
\frac{\sin \omega r_{12}}{4\pi\, \omega r_{12}} = \sum_{L=0}^{\infty} \sum_{M=-L}^{L} j_L(\omega r_<) j_L(\omega r_>) Y_{LM}^* (\hat{r}_1) Y_{LM} (\hat{r}_2).
\end{eqnarray}
Here $j_L$ is the spherical Bessel function of the first kind, $r_> = \max (\mathbf{r}_1, \mathbf{r}_2 )$ and $r_< = \min (\mathbf{r}_1, \mathbf{r}_2)$. Then, using the explicit form of the wave functions (\ref{wf_coord}), and performing the angular algebra, the following expression can be obtained:
\begin{eqnarray}
\label{M_el_expansion}
\langle ab|I( \omega )|cd \rangle = \sum_{JM} (-1)^{\varphi}
{\begin{pmatrix}j_a&J&j_c\\-m_{a}&M&m_{c} \end{pmatrix}}
\\
\nonumber
\times {\begin{pmatrix}j_b&J&j_d\\-m_{b}&-M&m_{d} \end{pmatrix}} 
\langle a b || I(\omega) || c d\rangle_J,
\end{eqnarray}
where $\varphi\equiv j_{a}-m_{a} + J - M + j_{b} - m_{b}$, and the reduced matrix element $\langle a b || I(\omega) || c d\rangle_J$ on the right-hand side of Eq.~(\ref{M_el_expansion}) is given in Appendix~\ref{ap:rml}, see Eq.~(\ref{M_el_f}). 

Substituting the expansion~(\ref{M_el_expansion}) into Eq.~(\ref{SE_general}) and summing over $m_n$ and $M$, one can obtain an expression convenient for numerical calculations:
\begin{eqnarray}
\label{se_mp}
\Delta \varepsilon_a^{\beta} = \frac{e^2}{\pi} \sum\limits_{J, n_{r_n}, \kappa_n} \frac{(-1)^{j_n-j_a+J}}{2 j_a+1} 
\\
\nonumber
\times \int\limits_{-\infty}^{\infty} d \omega \, n_\beta(|\omega|)   \frac{\langle a {n} || I(|\omega|) || {n}  a\rangle_J}{\varepsilon_{a} - \omega -\overline{\varepsilon}_n}. 
\end{eqnarray}

\subsection{Transitions between hyperfine sublevels}
\label{tr_hfs}
To account for the hyperfine structure of atomic levels, one can use a substitution
\begin{eqnarray}
\label{hfs-1}
|a\rangle \equiv |j_a m_a\rangle \rightarrow \sum_{M_I m_n} C^{F M_F}_{I M_I j_n m_n} |I M_I \rangle |j_n m_n\rangle,
\end{eqnarray}
where $I,M_I$ are the nuclear spin and its projection, $F$ is the total angular momentum of the atom, and $M_F$ denotes the projection of the moment $F$. 

Then, to calculate the thermal shift of the hyperfine sublevel, it is sufficient to consider the only contribution (closest in binding energy) in the sum of Eq.~(\ref{SE_general}) for the state $n$ of the same parity as $a$. For instance, for the ground state in $I=1/2$ hydrogen-like atom, we arrive at
\begin{eqnarray}
\label{SE_hfs}
\Delta \varepsilon_a^\beta = \frac{e^2}{\pi} \int\limits_{-\infty}^{\infty} d \omega \, n_\beta(|\omega|) 
\frac{ \langle 1s^{F=0} 1s^{F=1} | I(|\omega|) | 1s^{F=1} 1s^{F=0} \rangle}{\varepsilon_{1s}^{F=0} - \omega - \overline{\varepsilon}_{1s}^{F=1}}.
\end{eqnarray} 

After some angular algebra, one can obtain an expression more suitable for numerical calculations in this case
\begin{eqnarray}
\label{se_hfs_final}
\Delta \varepsilon_a^{\beta} = \frac{e^2}{\pi} \sum_{J} (-1)^{J}3 \left({\begin{Bmatrix}1&J&0\\1/2&1/2&1/2\end{Bmatrix}}\right)^2
\\
\nonumber
\times \int\limits_{-\infty}^{\infty} d \omega \, n_\beta(|\omega|)  \frac{\langle 1s 1s || I(|\omega|) || 1s 1s \rangle_J}{\Delta \varepsilon_{\rm HFS} - \omega + \mathrm{i} 0}
,
\end{eqnarray}
where $\Delta \varepsilon_{\rm HFS} = \varepsilon_{1s}^{F=0} - \varepsilon_{1s}^{F=1}$.


\section{Technical details, numerical results and discussion}

Below we evaluate the real part of the thermal one-loop self-energy correction numerically for excited states of the hydrogen atom and the ground state of various hydrogen-like ions. The Dirac equation is solved using the finite basis set method. The basis set is obtained from the expansion of the wave function in terms of B-splines within the framework of the dual-kinetic-balance (DKB) approach~\cite{DKB}. This method excludes the so-called spurious states and establishes the correct asymptotics of the wave functions in the non-relativistic limit. 

Within the DKB approach, the specific choice of basis functions corresponds to parameters such as the number of B-splines and the size of the box. By varying these values, one can control the convergence of numerical calculations and, with sufficiently short computation time, guarantee results accurate to approximately seven significant figures. It should also be noted that the DKB approach is directly applicable to the extended charge nucleus only. In further calculations, the spherical and Fermi models are used; the results do not depend on the choice of nuclear distribution. By replacing the full Dirac spectrum with a discrete pseudo-spectrum, summation over intermediate states presents no difficulty. Finally, integration over frequency is performed using Simpson quadrature, which also allows for effectively tracking convergence and the error of the numerical calculations.

\begin{figure}[t]
\centering
\includegraphics[width=0.8\columnwidth]{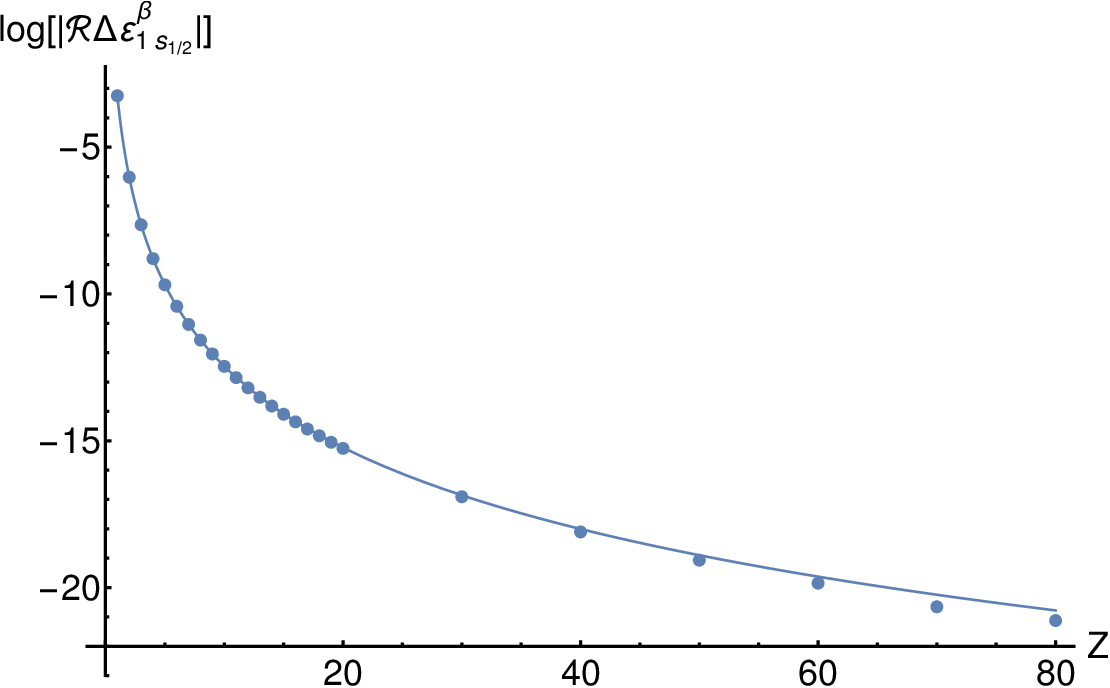}
\caption{Thermal shift for the ground state of hydrogen-like ions as a function of the nuclear charge $Z$ on a logarithmic scale. The solid line corresponds to the asymptotic estimate $1/Z^4$, with shift values in Hz taken from Table~\ref{table:re_1s_z_300k}.}
\label{fig:re_1s_z_300k}
\end{figure}

First, we present the results for the calculations of the energy shift of the ground state of hydrogen-like ions with the arbitrary nuclear charge $Z$. The real part of the thermal SE correction is calculated according to Eq.~(\ref{se_mp}). The obtained results for a temperature of $300$ K are listed in Table~\ref{table:re_1s_z_300k} and illustrated by the graph in Fig.~\ref{fig:re_1s_z_300k}, which shows the dependence of the energy shift on the nuclear charge $Z$. The parametric estimate for the real part of Eq.~(\ref{se_mp}) can be achieved in the standard way: $r\sim 1/(m\alpha Z)$, $\varepsilon\sim m(\alpha Z)^2$ in relativistic units. Then, in the dipole approximation, $I(|\omega|)\sim \bm{r}_1\bm{r}_2$, see~\cite{Lopez_2025}, combined with the static limit, one can find: $\Delta \varepsilon_a^{\beta} \sim (k_{\rm B}T)^4/m^3\alpha^3 Z^4$, i.e., $\Delta \varepsilon_a^{\beta}\sim 1/Z^4$.
\begin{table}[h!]
\centering
\caption{The real part of the one-loop self-energy correction for the $1s_{1/2}$ state of hydrogen-like ions in Hz at a temperature of $T=300$ K. For $Z < 10$, a spherical nuclear model is used, and for $Z \geq 10$, a Fermi nuclear model is used. Results from Ref.~\cite{farley} are marked with an asterisk $^*$, from Ref.~\cite{PRA_2015} with a cross $^\dagger$, and from Ref.~\cite{reex} with two asterisks $^{**}$. The estimate of the numerical calculation error is indicated in parentheses, and the order of the calculated shift is presented separately for convenience.
}
\begin{tabular}{p{2cm} l l} 
\hline\hline
$Z$ & $\Re \Delta \varepsilon_{1s_{1/2}}^{\beta}$, Hz & \\ 
\hline 
1	& -3.875176(6)  &      $\times 10^{-2}$        \\
	            & -4.128$^*$  &      $\times 10^{-2}$        \\
        & -3.87511$^\dagger$	  &      $\times 10^{-2}$        \\
	            & -3.88$^{**}$  &      $\times 10^{-2}$        \\
\hline 
2	& -2.42128(2)   &      $\times 10^{-3}$        \\
	            & -2.42$^{**}$  &      $\times 10^{-2}$        \\
\hline 
3	& -4.78121(7)   &      $\times 10^{-4}$        \\
4	& -1.51219(5)   &      $\times 10^{-4}$        \\
5	& -6.1898(2)    &      $\times 10^{-5}$        \\
6	& -2.9837(3)    &      $\times 10^{-5}$        \\
7	& -1.60918(10)  &      $\times 10^{-5}$        \\
8	& -9.4236(2)    &      $\times 10^{-6}$        \\
9	& -5.8781(8)    &      $\times 10^{-6}$        \\
10	& -3.8511(8)    &      $\times 10^{-6}$        \\
11	& -2.6268(4)    &      $\times 10^{-6}$        \\
12	& -1.8530(5)    &      $\times 10^{-6}$        \\
13	& -1.3424(2)    &      $\times 10^{-6}$        \\
14	& -9.974(4)     &      $\times 10^{-7}$        \\
15	& -7.543(2)     &      $\times 10^{-7}$        \\
16	& -5.824(2)     &      $\times 10^{-7}$        \\
17	& -4.560(2)     &      $\times 10^{-7}$        \\
18	& -3.626(2)     &      $\times 10^{-7}$        \\
19	& -2.904(3)     &      $\times 10^{-7}$        \\
20	& -2.360(2)     &      $\times 10^{-7}$        \\
30	& -4.543(10)    &      $\times 10^{-8}$        \\
40	& -1.370(3)     &      $\times 10^{-8}$        \\
50	& -5.24(8)      &      $\times 10^{-9}$        \\
60	& -2.39(2)      &      $\times 10^{-9}$        \\
70	& -1.07(2)      &      $\times 10^{-9}$        \\
80	& -6.7(2)       &      $\times 10^{-10}$       \\
\hline\hline
\end{tabular}
\label{table:re_1s_z_300k}
\end{table}




\textcolor{black}{
The numerical calculation error given in parentheses in Table I is determined as follows. First, the partial-wave decomposition in Eq.~(\ref{PE}) constitutes a rapidly convergent series owing to the small value of the parameter $\omega$, which is defined by the Planck distribution (temperature). Extrapolation of the partial-wave series is not required and this effect can therefore be disregarded when estimating the calculation error. We then performed calculations with different sets of basis functions to determine the convergence of the result. The stated mean value was chosen considering a sufficiently fast computation time while ensuring a convergent result with the number of basis functions set to $140$ B-splines, with convergence already observed at $100$ B‑splines. Another parameter is the box size (was finally set to $600$ atomic units), which allows one to control the number of recovered 'real' atomic bound states. Recall that the DKB method replaces the atomic spectrum with an effective, purely discrete set of states. Because the continuous spectrum mixes with the discrete energy levels in this case, only the lowest few states correspond well to the 'real' ones. Their number depends on the box size and can be determined by comparing the Dirac energy and the pseudolevel energy. Thus, by varying these two parameters, we have, in our view, achieved optimal convergence of the obtained values with the corresponding stated numerical error. The same consideration regarding the pseudo-spectrum results in a larger numerical error for highly excited states (see below). We also note that, in practice, all numerical calculations (those presented above and subsequently) were carried out to machine precision.
}

The results from the Table~\ref{table:re_1s_z_300k} allow us to unambiguously track the effectiveness of the presented estimate for the thermal shift Eq.~(\ref{se_mp}), compare, for example, $Z=1$ and $Z=10$. It is obvious that the estimate derived for the dipole (long-wavelength) approximation begins to perform less effectively with an increase in the nuclear charge $Z$. For greater clarity, a fitting curve of $1/Z^4$ is plotted in the Fig.~\ref{fig:re_1s_z_300k} as a solid curve, on which the deviation from the estimate becomes noticeable starting from approximately $Z=30$.

Despite the fact that the numerical calculation worsens with an increasing nuclear charge, the deviation from the asymptotic behavior of $1/Z^4$ can be explained by relativistic corrections to the wave functions and the self-energy operator, as well as higher multipole moments \cite{Lopez_2025}. The estimate $\Delta \varepsilon_a^{\beta}$ for large $Z$ confirms that the thermal shift becomes negligible in highly relativistic hydrogen-like systems. Based on this, we focus mainly on light hydrogen-like systems.




The Table~\ref{table:re_exc_h_300k} indicates the real part of the self-energy correction for the excited states of hydrogen $\Re \Delta \varepsilon_{a}^{\beta}$. Within the framework of our fully relativistic method, the values for the thermal dynamic effect Eqs.~(\ref{SE_general}), (\ref{op:I}) align with good accuracy to the Stark shift obtained using a purely quantum mechanical approach \cite{farley}. The most significant discrepancy (around $8\%$ for the $2s$ state) arises from the fine structure of the levels, as well as from the Lamb shift and the numerical calculation method, which did not allow for the determination of as many significant figures as indicated in Table~\ref{table:re_exc_h_300k}; these factors were not accounted for in \cite{farley,PRA_2015,jetp2022}. In turn, comparison with \cite{reex} shows excellent agreement of the results. We should emphasize that the accuracy of our results can be confirmed by calculations of the thermal Zeeman shift (see below) without additional analytical derivations \cite{Lopez_2025} and carefully accounts for the fine structure within a fully relativistic approach (in contrast to \cite{jetp2022}). 
\begin{widetext}
\begin{center}
\begin{table}[ht!]
\caption{Real part of the self-energy correction for the excited states of hydrogen $\Re \Delta \varepsilon_{a}^{\beta}$ in Hz. The temperature of $300$ K is considered. Results with included Lamb shift are highlighted in italics. Numbers from Ref.~\cite{farley} are marked with an asterisk $^*$, from Ref.~\cite{PRA_2015} with a dagger $^\dagger$, and from Ref.~\cite{jetp2022} with two asterisks $^{**}$. }
\begin{tabular}{p{.5cm}p{2.2cm}p{2.2cm}p{2.2cm}p{2.2cm}p{2.2cm}p{2.2cm}p{2.2cm}}
\hline\hline
n & $s_{1/2}$ & $p_{1/2}$ & $p_{3/2}$ & $d_{3/2}$ & $d_{5/2}$ & $f_{5/2}$ & $f_{7/2}$ \\ 
\hline
2& $-0.98807(2)$      & $-1.5203366(2)$  &  $-1.54457(1)$  & & & & \\                                                                                      
  & ${\it-0.99041(1)}$ & ${\it-1.5197110(1)}$ & ${\it-1.54408(1)}$ & & \\ 
   & $-0.989702^\dagger$   & & & & & & \\                                                                                                                  
& $-1.04^{**}$ & & & & & & \\
   &  $-1.077^*$    &   $-1.535^*$  &   $-1.535^*$  & & & & \\                                                                                             
\hline
3& $-8.866970(4)$       & $-11.416930(2)$   &  $-11.498190(1)$  & $-16.548507(1)$   & $-16.5325918(8)$   & & \\                                            
  & ${\it-8.866014(3)}$ & ${\it-11.417380(1)}$ & ${\it-11.498705(1)}$ & ${\it-16.548497(1)}$ & ${\it-16.5326016(4)}$ & \\ 
& $-8.93974^\dagger$    & & & & & & \\    
   & $-8.79^{**}$ & & & $-16.35^{**}$ & $-16.35^{**}$ & & \\
   &  $-9.103^*$    &   $-11.51^*$  &   $-11.51^*$  &   $-16.60^*$  &   $-16.60^*$  & & \\                                                                 
\hline
4& $-50.656666(5)$        & $-60.03414(6)$    &  $-60.1675(2)$   & $-79.0932(1)$     & $-79.0740(1)$     & $-108.0170(1)$    & $-108.01235(8)$   \\         
  & ${\it-50.653648(9)}$ & ${\it-60.03540(1)}$ & ${\it-60.1687(1)}$ & ${\it-79.0933(1)}$ & ${\it-79.0740(1)}$ & ${\it-108.0170(2)}$ & ${\it-108.01241(2)}$ \\ 
& $-50.1879^\dagger$    & & & & & & \\ 
   & $-50.80^{**}$ & & & $-78.39^{**}$ & $-78.39^{**}$ & & \\
   &  $-51.19^*$    &   $-60.37^*$  &   $-60.37^*$  &   $-79.36^*$  &   $-79.36^*$  &   $-108.15^*$ &   $-108.15^*$ \\                                     
\hline
5& $-208.460(3)$        & $-234.852(4)$     &  $-235.0491(3)$    & $-288.257(1)$     & $-288.229(6)$     & $-369.093(4)$     & $-369.085(2)$     \\        
  & ${\it-208.468(1)}$ & ${\it-234.872(1)}$ & ${\it-235.066(1)}$ & ${\it-288.281(2)}$ & ${\it-288.258(2)}$ & ${\it-369.130(3)}$ & ${\it-369.125(3)}$ \\ 
& $-186.884^\dagger$    & & & & & & \\      
   &  $-209.5^*$    &   $-235.6^*$  &   $-235.6^*$  &   $-288.8^*$  &   $-288.8^*$  &   $-369.37^*$ &   $-369.37^*$ \\                                     
\hline
6& $-273.12(2)$        & $-290.308(9)$       &  $-290.96(1)$   & $-322.10(2)$       & $-321.08(4)$       & $-362.02(1)$       & $-362.02(4)$      \\       
  & ${\it-273.16(1)}$ & ${\it-290.406(2)}$ & ${\it-290.27(1)}$ & ${\it-322.24(1)}$ & ${\it-321.60(1)}$ & ${\it-362.25(1)}$ & ${\it-362.23(1)}$ \\ 
   & $-273.48^{**}$ & & & $-318.86^{**}$ & $-318.86^{**}$ & & \\  
   &  $-274.7^*$    &   $-291.5^*$  &   $-291.5^*$  &   $-323.0^*$  &   $-323.0^*$  &   $-362.27^*$ &   $-362.27^*$ \\                                     
\hline
7& $4.060(4)$         & $17.88(8)$       &  $17.6(1)$      & $50.4(1)$       & $50.5(1)$       & $112.36(9)$     & $112.31(9)$      \\                     
  & ${\it4.44(1)}$ & ${\it17.96(1)}$ & ${\it18.11(1)}$ & ${\it50.49(1)}$ & ${\it51.01(1)}$ & ${\it112.38(1)}$ & ${\it112.98(1)}$ \\ 
   &  $1.344^*$     &   $15.60^*$   &   $15.60^*$   &   $48.57^*$   &   $48.57^*$   &   $111.54^*$  &   $111.54^*$ \\                                      
\hline
8& $398.61(6)$           & $429.12(7)$        &  $428.93(8)$       & $493.65(9)$        & $494.1(1)$        & $601.15(9)$        & $601.56(8)$       \\    
  & ${\it398.80(1)}$ & ${\it429.35(1)}$ & ${\it429.07(1)}$ & ${\it494.23(1)}$ & ${\it494.22(1)}$ & ${\it601.52(1)}$ & ${\it601.69(2)}$ \\ 
   & $398.08^{**}$ & & & $505.91^{**}$ & $505.91^{**}$ & & \\  
   &  $393.9^*$     &   $424.8^*$   &   $424.8^*$   &   $490.3^*$   &   $490.3^*$   &   $599.07^*$  &   $599.07^*$ \\                                      
\hline
9& $766.72(4)$         & $802.03(5)$        &  $801.61(2)$    & $875.32(7)$        & $875.42(3)$        & $990.3(1)$        & $990.33(5)$       \\         
  & ${\it767.05(1)}$ & ${\it802.18(1)}$ & ${\it802.24(3)}$ & ${\it875.33(1)}$ & ${\it875.61(2)}$ & ${\it990.24(1)}$ & ${\it990.29(1)}$ \\ 
   &  $761.1^*$     &   $797.1^*$   &   $797.1^*$   &   $870.9^*$   &   $870.9^*$   &   $987.45^*$  &   $987.45^*$ \\                                      
\hline
10& $1074.73(1)$       & $1110.11(3)$     &  $1109.70(4)$    & $1181.13(8)$    & $1180.92(4)$     & $1288.86(9)$     & $1288.71(2)$     \\                 
  & ${\it1075.05(1)}$ & ${\it1109.93(1)}$ & ${\it1109.85(1)}$ & ${\it1180.84(2)}$ & ${\it1180.84(2)}$ & ${\it1288.90(1)}$ & ${\it1288.58(2)}$ \\ 
  &   $1073^*$      &   $1108^*$    &   $1108^*$    &   $1180^*$    &   $1180^*$    &   $1289^*$    &   $1289^*$ \\                                        
\hline                                                                                                                                                     
11& $1322.72(3)$         & $1355.58(2)$       &  $1355.38(2)$      & $1420.16(2)$       & $1420.352(5)$    & $1515.74(2)$       & $1515.90(4)$      \\     
  & ${\it1323.30(2)}$ & ${\it1355.42(2)}$ & ${\it1355.19(1)}$ & ${\it1419.83(1)}$ & ${\it1419.83(1)}$ & ${\it1515.77(1)}$ & ${\it1515.71(1)}$ \\ 
  &   $1327^*$      &   $1360^*$    &   $1360^*$    &   $1425^*$    &   $1425^*$    &   $1523^*$    &   $1523^*$ \\                                        
\hline                                                                                                                                                     
12& $1520.16(1)$       & $1548.95(2)$  &  $1550.20(3)$  & $1605.77(8)$     & $1605.88(4)$     & $1689.097(8)$       & $1689.28(3)$    \\                   
  & ${\it1520.79(1)}$ & ${\it1548.89(1)}$ & ${\it1548.63(1)}$ & ${\it1605.87(2)}$ & ${\it1605.82(1)}$ & ${\it1689.070(2)}$ & ${\it1689.17(1)}$ \\ 
   & $1520.74^{**}$ & & & $1638.03^{**}$ & $1638.03^{**}$ & & \\                                                                                           
  &   $1533^*$      &   $1563^*$    &   $1563^*$    &   $1620^*$    &   $1620^*$    &   $1705^*$    &   $1705^*$ \\          
\hline\hline
\end{tabular}
\label{table:re_exc_h_300k}
\end{table} 
\end{center}
\end{widetext}

Specifically, the values given in the Table~\ref{table:re_exc_h_300k} allow for an unambiguous determination of the influence of the Lamb shift on the thermal shifts of hydrogen atomic levels. It can be seen that the contribution does not exceed a tenth of a percent of the total shift. In turn, the values show a fundamental difference between two approaches: a fully relativistic one that takes into account the Lamb shift (this work), and an approximate non-relativistic \cite{farley}.

\textcolor{black}{
In addition to Table~\ref{table:re_exc_h_300k}, we have performed calculations of the thermal one-loop correction for several hydrogen-like ions at various temperatures for low-lying states. The results are collected in Table~\ref{tab:03}.}
\begin{widetext}
\begin{center}
\begin{table}[ht!]
\caption{\textcolor{black}{The real part of the thermal one-loop correction Eq.~(\ref{se_mp}) for the low-lying states of hydrogen and hydrogen-like ions. Values are given in Hz for various temperatures (in Kelvin). Results with included Lamb shift are highlighted in italics. As before, the estimated numerical error is indicated in parentheses. The corresponding one-electron atomic system is specified by the nuclear charge $Z$.}
}
\resizebox{\textwidth}{!}{

\begin{tabular}{l c c c c c c c c c} 
\hline\hline
$Z \quad$ & State & $3$ K & $5$ K & $10$ K & $20$ K & $30$ K & $50$ K & $100$ K & $200$ K \\ 
\hline
$ 1 $ & $1s_{1/2}$
& $ -4.3(4)\times 10^{-10} $ & $ -3.1(1)\times 10^{-9} $ & $ -4.83(5)\times 10^{-8} $ & $ -7.67(2)\times 10^{-7} $ & $ -3.879(4)\times 10^{-6} $ & $ -2.991(1)\times 10^{-5} $ & $ -4.7842(5)\times 10^{-4} $ & $ -7.6545(2)\times 10^{-3} $ \\
& $2s_{1/2}$
& $ 4.8(1)\times 10^{-6} $ & $ 1.24(1)\times 10^{-5} $ & $ 5.08(2)\times 10^{-5} $ & $ 1.897(5)\times 10^{-4} $ & $ 3.724(8)\times 10^{-4} $ & $ 5.32(1)\times 10^{-4} $ & $ -7.416(5)\times 10^{-3} $ & $ -1.82900(3)\times 10^{-1} $ \\
 &  & ${\it4.58(6)\times 10^{-6}}$ & ${\it1.20(3)\times 10^{-5}}$ & ${\it4.93(5)\times 10^{-5}}$ & ${\it1.747(8)\times 10^{-4}}$ & ${\it3.42(1)\times 10^{-4}}$ & ${\it4.82(3)\times 10^{-4}}$ & ${\it-7.637(3)\times 10^{-3}}$ & ${\it-1.84018(8)\times 10^{-1}}$\\
& $2p_{1/2}$
& $ -1.52(2)\times 10^{-8} $ & $ -1.171(7)\times 10^{-7} $ & $ -1.872(3)\times 10^{-6} $ & $ -2.994(1)\times 10^{-5} $ & $ -1.5156(2)\times 10^{-4} $ & $ -1.16946(7)\times 10^{-3} $ & $ -1.87161(3)\times 10^{-2} $ & $ -2.99777(1)\times 10^{-1} $ \\
 &  & ${\it4.78(2)\times 10^{-8}}$ & ${\it5.76(7)\times 10^{-8}}$ & ${\it-1.175(3)\times 10^{-6}}$ & ${\it-2.716(1)\times 10^{-5}}$ & ${\it-1.4530(2)\times 10^{-4}}$ & ${\it-1.15208(7)\times 10^{-3}}$ & ${\it-1.86466(3)\times 10^{-2}}$ & ${\it-2.99499(1)\times 10^{-1}}$\\
& $2p_{3/2}$
& $ -2.42(7)\times 10^{-6} $ & $ -6.35(6)\times 10^{-6} $ & $ -2.79(1)\times 10^{-5} $ & $ -1.350(2)\times 10^{-4} $ & $ -3.894(4)\times 10^{-4} $ & $ -1.8340(7)\times 10^{-3} $ & $ -2.1389(2)\times 10^{-2} $ & $ -3.10529(1)\times 10^{-1} $ \\
 &  & ${\it-2.38(3)\times 10^{-6}}$ & ${\it-6.3(2)\times 10^{-6}}$ & ${\it-2.79(3)\times 10^{-5}}$ & ${\it-1.305(4)\times 10^{-4}}$ & ${\it-3.810(7)\times 10^{-4}}$ & ${\it-1.828(1)\times 10^{-3}}$ & ${\it-2.1354(1)\times 10^{-2}}$ & ${\it-3.10272(4)\times 10^{-1}}$\\
\hline
$ 5 $ & $1s_{1/2}$
& $ 1.3\times 10^{-12} $ & $ 5.6\times 10^{-13} $ & $ -5.5\times 10^{-11} $ & $ -1.1(4)\times 10^{-9} $ & $ -6.0(8)\times 10^{-9} $ & $ -4.7(2)\times 10^{-8} $ & $ -7.62(9)\times 10^{-7} $ & $ -1.222(4)\times 10^{-5} $ \\
& $2s_{1/2}$
& $ -4.610(2)\times 10^{-8} $ & $ -3.2532(5)\times 10^{-7} $ & $ -5.0616(2)\times 10^{-6} $ & $ -8.54177(9)\times 10^{-5} $ & $ -4.94906(2)\times 10^{-4} $ & $ -4.249908(6)\times 10^{-3} $ & $ -1.48846(1)\times 10^{-2} $ & $ 1.65475(2)\times 10^{-1} $ \\
 &  & ${\it-9.149(2)\times 10^{-8}}$ & ${\it-4.5153(5)\times 10^{-7}}$ & ${\it-5.5688(2)\times 10^{-6}}$ & ${\it-8.74875(9)\times 10^{-5}}$ & ${\it-4.99773(2)\times 10^{-4}}$ & ${\it-4.264815(6)\times 10^{-3}}$ & ${\it-1.49319(1)\times 10^{-2}}$ & ${\it1.65378(4)\times 10^{-1}}$\\
& $2p_{1/2}$
& $ 6.305(9)\times 10^{-9} $ & $ 1.738(2)\times 10^{-8} $ & $ 6.719(10)\times 10^{-8} $ & $ 2.320(4)\times 10^{-7} $ & $ 3.838(8)\times 10^{-7} $ & $ -1.65(2)\times 10^{-7} $ & $ -2.3045(9)\times 10^{-5} $ & $ -4.4800(4)\times 10^{-4} $ \\
 &  & ${\it8.81(2)\times 10^{-9}}$ & ${\it2.434(2)\times 10^{-8}}$ & ${\it9.487(9)\times 10^{-8}}$ & ${\it3.433(4)\times 10^{-7}}$ & ${\it6.338(9)\times 10^{-7}}$ & ${\it5.30(2)\times 10^{-7}}$ & ${\it-2.0267(9)\times 10^{-5}}$ & ${\it-4.3689(4)\times 10^{-4}}$\\
& $2p_{3/2}$
& $ 1.9848(9)\times 10^{-8} $ & $ 1.5361(2)\times 10^{-7} $ & $ 2.49167(10)\times 10^{-6} $ & $ 4.25045(4)\times 10^{-5} $ & $ 2.468149(9)\times 10^{-4} $ & $ 2.121595(3)\times 10^{-3} $ & $ 7.39879(6)\times 10^{-3} $ & $ -8.33945(10)\times 10^{-2} $ \\
 &  & ${\it3.9834(9)\times 10^{-8}}$ & ${\it2.0919(2)\times 10^{-7}}$ & ${\it2.71512(10)\times 10^{-6}}$ & ${\it4.34194(4)\times 10^{-5}}$ & ${\it2.4906(3)\times 10^{-4}}$ & ${\it2.135(2)\times 10^{-3}}$ & ${\it7.60(7)\times 10^{-3}}$ & ${\it-8.22(4)\times 10^{-2}}$\\
\hline
$ 10 $ & $1s_{1/2}$
& $ 2.8\times 10^{-12} $ & $ 7.5\times 10^{-12} $ & $ 2.7\times 10^{-11} $ & $ 4.9\times 10^{-11} $ & $ -1.0\times 10^{-10} $ & $ -2(1)\times 10^{-9} $ & $ -4.4(6)\times 10^{-8} $ & $ -7.5(2)\times 10^{-7} $ \\
& $2s_{1/2}$
& $ -3.860(5)\times 10^{-8} $ & $ -1.1441(9)\times 10^{-7} $ & $ -5.236(1)\times 10^{-7} $ & $ -3.0224(4)\times 10^{-6} $ & $ -1.02470(10)\times 10^{-5} $ & $ -5.9106(3)\times 10^{-5} $ & $ -8.1443(1)\times 10^{-4} $ & $ -1.275844(4)\times 10^{-2} $ \\
 &  & ${\it-4.998(5)\times 10^{-8}}$ & ${\it-1.4591(9)\times 10^{-7}}$ & ${\it-6.494(1)\times 10^{-7}}$ & ${\it-3.5253(4)\times 10^{-6}}$ & ${\it-1.13786(10)\times 10^{-5}}$ & ${\it-6.2250(3)\times 10^{-5}}$ & ${\it-8.2703(1)\times 10^{-4}}$ & ${\it-1.280920(4)\times 10^{-2}}$\\
& $2p_{1/2}$
& $ 3.799(1)\times 10^{-8} $ & $ 1.0965(3)\times 10^{-7} $ & $ 4.4716(6)\times 10^{-7} $ & $ 1.79706(2)\times 10^{-6} $ & $ 4.03993(4)\times 10^{-6} $ & $ 1.11540(1)\times 10^{-5} $ & $ 4.32084(5)\times 10^{-5} $ & $ 1.50089(2)\times 10^{-4} $ \\
 &  & ${\it3.854(6)\times 10^{-8}}$ & ${\it1.116(4)\times 10^{-7}}$ & ${\it4.541(2)\times 10^{-7}}$ & ${\it1.8226(4)\times 10^{-6}}$ & ${\it4.1076(1)\times 10^{-6}}$ & ${\it1.1316(7)\times 10^{-5}}$ & ${\it4.3896(9)\times 10^{-5}}$ & ${\it1.5280(3)\times 10^{-4}}$\\
& $2p_{3/2}$
& $ 3.07(7)\times 10^{-10} $ & $ 2.37(2)\times 10^{-9} $ & $ 3.796(8)\times 10^{-8} $ & $ 6.075(3)\times 10^{-7} $ & $ 3.0764(7)\times 10^{-6} $ & $ 2.3763(2)\times 10^{-5} $ & $ 3.82188(8)\times 10^{-4} $ & $ 6.24933(3)\times 10^{-3} $ \\
 &  & ${\it5.292(7)\times 10^{-9}}$ & ${\it1.622(2)\times 10^{-8}}$ & ${\it9.334(8)\times 10^{-8}}$ & ${\it8.290(3)\times 10^{-7}}$ & ${\it3.5749(7)\times 10^{-6}}$ & ${\it2.5149(2)\times 10^{-5}}$ & ${\it3.87740(8)\times 10^{-4}}$ & ${\it6.27172(3)\times 10^{-3}}$\\
\hline
$ 30 $ & $1s_{1/2}$
& $ 2.4\times 10^{-12} $ & $ 6.6\times 10^{-12} $ & $ 2.6\times 10^{-11} $ & $ 1.0\times 10^{-10} $ & $ 2.3\times 10^{-10} $ & $ 6.2\times 10^{-10} $ & $ 2.1\times 10^{-9} $ & $ 1.6\times 10^{-9} $ \\
& $2s_{1/2}$
& $ 1.06(1)\times 10^{-9} $ & $ 8.2(5)\times 10^{-9} $ & $ 1.397(1)\times 10^{-7} $ & $ 2.8793(3)\times 10^{-6} $ & $ 1.2856(1)\times 10^{-5} $ & $ 2.6091(7)\times 10^{-5} $ & $ -2.7323(3)\times 10^{-4} $ & $ -2.52239(2)\times 10^{-3} $ \\
 &  & ${\it-6.6(1)\times 10^{-10}}$ & ${\it3.4(5)\times 10^{-9}}$ & ${\it1.205(1)\times 10^{-7}}$ & ${\it2.7992(3)\times 10^{-6}}$ & ${\it1.2673(1)\times 10^{-5}}$ & ${\it2.5623(7)\times 10^{-5}}$ & ${\it-2.7479(2)\times 10^{-4}}$ & ${\it-2.52805(3)\times 10^{-3}}$\\
& $2p_{1/2}$
& $ -1.051(3)\times 10^{-9} $ & $ -8.2(2)\times 10^{-9} $ & $ -1.398(7)\times 10^{-7} $ & $ -2.8807(3)\times 10^{-6} $ & $ -1.28638(3)\times 10^{-5} $ & $ -2.6155(3)\times 10^{-5} $ & $ 2.72201(10)\times 10^{-4} $ & $ 2.50577(5)\times 10^{-3} $ \\
 &  & ${\it-4.87(4)\times 10^{-10}}$ & ${\it-6.6(2)\times 10^{-9}}$ & ${\it-1.333(7)\times 10^{-7}}$ & ${\it-2.871(3)\times 10^{-6}}$ & ${\it-1.234(1)\times 10^{-5}}$ & ${\it-3.2(3)\times 10^{-5}}$ & ${\it2.6(1)\times 10^{-4}}$ & ${\it2.33(8)\times 10^{-3}}$\\
& $2p_{3/2}$
& $ 1.9\times 10^{-12} $ & $ 7.2\times 10^{-12} $ & $ 6(4)\times 10^{-11} $ & $ 8(2)\times 10^{-10} $ & $ 4.0(4)\times 10^{-9} $ & $ 3.0(1)\times 10^{-8} $ & $ 4.81(4)\times 10^{-7} $ & $ 7.68(2)\times 10^{-6} $ \\
 &  & ${\it5.45(4)\times 10^{-10}}$ & ${\it1.52(1)\times 10^{-9}}$ & ${\it6.10(4)\times 10^{-9}}$ & ${\it2.50(2)\times 10^{-8}}$ & ${\it5.83(4)\times 10^{-8}}$ & ${\it1.81(1)\times 10^{-7}}$ & ${\it1.085(4)\times 10^{-6}}$ & ${\it1.009(2)\times 10^{-5}}$\\
\hline
$ 60 $ & $1s_{1/2}$
& $ 1.6\times 10^{-12} $ & $ 4.4\times 10^{-12} $ & $ 1.8\times 10^{-11} $ & $ 7.0\times 10^{-11} $ & $ 1.6\times 10^{-10} $ & $ 4.4\times 10^{-10} $ & $ 1.7\times 10^{-9} $ & $ 6.6\times 10^{-9} $ \\
& $2s_{1/2}$
& $ 8(8)\times 10^{-12} $ & $ 4(2)\times 10^{-11} $ & $ 5.0(9)\times 10^{-10} $ & $ 7.5(3)\times 10^{-9} $ & $ 3.73(8)\times 10^{-8} $ & $ 2.85(2)\times 10^{-7} $ & $ 4.557(8)\times 10^{-6} $ & $ 7.324(3)\times 10^{-5} $ \\
 &  & ${\it-3.78(8)\times 10^{-10}}$ & ${\it-1.03(2)\times 10^{-9}}$ & ${\it-3.79(9)\times 10^{-9}}$ & ${\it-9.7(3)\times 10^{-9}}$ & ${\it-1.4(8)\times 10^{-9}}$ & ${\it1.78(2)\times 10^{-7}}$ & ${\it4.127(8)\times 10^{-6}}$ & ${\it7.152(3)\times 10^{-5}}$\\
& $2p_{1/2}$
& $ -2.4\times 10^{-12} $ & $ -2.5(9)\times 10^{-11} $ & $ -4.4(3)\times 10^{-10} $ & $ -7.2(1)\times 10^{-9} $ & $ -3.68(3)\times 10^{-8} $ & $ -2.847(9)\times 10^{-7} $ & $ -4.565(4)\times 10^{-6} $ & $ -7.346(2)\times 10^{-5} $ \\
 &  & ${\it1.22(3)\times 10^{-10}}$ & ${\it3.20(9)\times 10^{-10}}$ & ${\it9.4(3)\times 10^{-10}}$ & ${\it-1.7(1)\times 10^{-9}}$ & ${\it-2.44(3)\times 10^{-8}}$ & ${\it-2.502(9)\times 10^{-7}}$ & ${\it-4.428(4)\times 10^{-6}}$ & ${\it-7.291(2)\times 10^{-5}}$\\
& $2p_{3/2}$
& $ 1.5\times 10^{-12} $ & $ 4.2\times 10^{-12} $ & $ 1.7\times 10^{-11} $ & $ 7.5\times 10^{-11} $ & $ 1.9\times 10^{-10} $ & $ 7.5\times 10^{-10} $ & $ 7(3)\times 10^{-9} $ & $ 9(1)\times 10^{-8} $ \\
 &  & ${\it1.28(3)\times 10^{-10}}$ & ${\it3.56(9)\times 10^{-10}}$ & ${\it1.42(3)\times 10^{-9}}$ & ${\it5.7(1)\times 10^{-9}}$ & ${\it1.28(3)\times 10^{-8}}$ & ${\it3.59(9)\times 10^{-8}}$ & ${\it1.48(3)\times 10^{-7}}$ & ${\it6.5(1)\times 10^{-7}}$\\
\hline
$ 92 $ & $1s_{1/2}$
& $ -1.6\times 10^{-11} $ & $ -4.5\times 10^{-11} $ & $ -1.8\times 10^{-10} $ & $ -7.1\times 10^{-10} $ & $ -1.6\times 10^{-9} $ & $ -4.5\times 10^{-9} $ & $ -1.8\times 10^{-8} $ & $ -7.1\times 10^{-8} $ \\
& $2s_{1/2}$
& $ 5.4\times 10^{-14} $ & $ 3.0\times 10^{-13} $ & $ 4.0\times 10^{-12} $ & $ 6.2\times 10^{-11} $ & $ 3.1\times 10^{-10} $ & $ 2.4\times 10^{-9} $ & $ 4(3)\times 10^{-8} $ & $ 6(1)\times 10^{-7} $ \\
 &  & ${\it-1.3(2)\times 10^{-10}}$ & ${\it-3.6(7)\times 10^{-10}}$ & ${\it-1.4(3)\times 10^{-9}}$ & ${\it-6(1)\times 10^{-9}}$ & ${\it-1.3(2)\times 10^{-8}}$ & ${\it-3.4(7)\times 10^{-8}}$ & ${\it-1.1(3)\times 10^{-7}}$ & ${\it2.4\times 10^{-8}}$\\
& $2p_{1/2}$
& $ 4.3\times 10^{-13} $ & $ 1.0\times 10^{-12} $ & $ 1.3\times 10^{-12} $ & $ -4.1\times 10^{-11} $ & $ -3(1)\times 10^{-10} $ & $ -2.3(3)\times 10^{-9} $ & $ -3.8(1)\times 10^{-8} $ & $ -6.18(5)\times 10^{-7} $ \\
 &  & ${\it4.0(1)\times 10^{-11}}$ & ${\it1.12(3)\times 10^{-10}}$ & ${\it4.4(1)\times 10^{-10}}$ & ${\it1.73(5)\times 10^{-9}}$ & ${\it3.7(1)\times 10^{-9}}$ & ${\it8.8(3)\times 10^{-9}}$ & ${\it6(1)\times 10^{-9}}$ & ${\it-4.41(5)\times 10^{-7}}$\\
& $2p_{3/2}$
& $ 1.2\times 10^{-12} $ & $ 3.5\times 10^{-12} $ & $ 1.4\times 10^{-11} $ & $ 5.6\times 10^{-11} $ & $ 1.3\times 10^{-10} $ & $ 3.5\times 10^{-10} $ & $ 1.5\times 10^{-9} $ & $ 7.7\times 10^{-9} $ \\
 &  & ${\it4.7(3)\times 10^{-11}}$ & ${\it1.32(8)\times 10^{-10}}$ & ${\it5.3(3)\times 10^{-10}}$ & ${\it2.1(1)\times 10^{-9}}$ & ${\it4.7(3)\times 10^{-9}}$ & ${\it1.32(8)\times 10^{-8}}$ & ${\it5.3(3)\times 10^{-8}}$ & ${\it2.1(1)\times 10^{-7}}$\\
\hline\hline
\end{tabular}
}
\label{tab:03}
\end{table}
\end{center}
\end{widetext}

\textcolor{black}{
Several observations can be drawn from Table~\ref{tab:03}. Considering the ground state of the hydrogen atom, a clear $T^4$ dependence is evident. This can be verified by comparing the values at temperatures of $10$ K and $100$ K, or $20$ K and $200$ K. This dependence holds approximately up to $Z = 10$, beyond which a deviation from this asymptotic behavior is observed. For $Z = 30$ and higher, the $T^4$ dependence is replaced by a quadratic one, which is explicitly demonstrated for $Z = 92$. We attribute this to the fact that the $T^4$ estimate was derived within the non-relativistic limit, which, as the results in Table~\ref{tab:03} demonstrate, is no longer sufficiently accurate for $Z > 10$. Thus, relativistic thermal effects become fundamentally important for high nuclear charges.
}

\textcolor{black}{
Another equally interesting effect is the deviation from the $Z^{-4}$ dependence with increasing nuclear charge at low temperatures. The breakdown of the non-relativistic $Z^{-4}$ scaling occurs earlier at temperatures below room temperature. For instance, this is clearly manifested for the ground state at temperatures of $50$ K and $10$ K for $Z = 1$ and $Z = 10$, respectively. We also note the sign-alternating behavior of the thermal shift with varying temperature and nuclear charge. Again, since this estimate was obtained in the non-relativistic limit, we attribute this behavior to the influence of relativistic effects. A discussion of the latter can be found in \cite{Lopez_2025}.
}

Being the most accurately calculated to date for the hydrogen atom, the values for the thermal shift, presented in Table~\ref{table:re_exc_h_300k}, can be used to determine the thermal frequency shifts in transitions between fine 
sublevels. In accordance with the quantum mechanical formalism for incorporating thermal radiation corrections to atomic levels \cite{farley}, or via the QED one-loop perturbative framework \cite{PRA_2015,Lopez_2025}, the dominant thermal contribution arises due to the static dipole polarizability. Considering that in this case the perturbation operator does not act on the spin part of the electronic function, such contributions cancel out for states with equal principal quantum number $n$ and orbital angular momentum $l$, but different total angular momenta $j$. 

The detailed theoretical analysis assigning the influence of the "electric component"\, (Stark effect) and the "magnetic component"\, (Zeeman effect), as well as the derivation of other thermal corrections, has been presented quite recently in the work \cite{Lopez_2025}. A distinctive feature of the approach presented in this study is that the fine structure of atomic levels is automatically taken into account within the framework of Dirac's theory. Thus, by performing calculations with a sufficient number of significant digits, we can accurately account for the mentioned reduction. 

The frequency shift for transitions between sublevels of fine structure can be determined by subtracting the corresponding values from Table~\ref{table:re_exc_h_300k}. For convenience, the values of the thermal shift (Zeeman plus relativistic effects) for the fine transition frequencies are collected in Table~\ref{tab:3}.
\begin{table}[htbp!]
\centering
\caption{The real part of the thermal one-loop SE correction Eq.~(\ref{se_mp}) to the frequencies of the lowest transitions between fine structure states in the hydrogen atom. Values are given in Hz at a temperature of $T=300$ K. The estimated numerical error is indicated in parentheses, as before.
}
\resizebox{\columnwidth}{!}{
\begin{tabular}{l c || l c} 
\hline\hline
Transition & $\Re \Delta \varepsilon_a^{\beta}-\Re \Delta \varepsilon_{a'}^{\beta}$ & Transition & $\Re \Delta \varepsilon_a^{\beta}-\Re \Delta \varepsilon_{a'}^{\beta}$ \\ 
\hline 
$2p_{3/2}-2p_{1/2}$	& $-0.02423(1)$  &      $3d_{5/2}-3d_{3/2}$   &  $0.01591(1)$   \\
$3p_{3/2}-3p_{1/2}$	& $-0.08126(2)$  &      $4d_{5/2}-4d_{3/2}$   & $0.0192(1)$   \\
$4p_{3/2}-4p_{1/2}$	& $-0.1331(1)$  &      $5d_{5/2}-5d_{3/2}$   &  $0.028(6)$  \\
$5p_{3/2}-5p_{1/2}$	& $-0.197(4)$  &      $4f_{7/2}-4f_{5/2}$   &  $0.0046(8)$  \\
$6p_{3/2}-6p_{1/2}$	& $-0.65(1)$  &      $5f_{7/2}-5f_{5/2}$   &  $0.008(4)$  \\
\hline\hline
\end{tabular}
}
\label{tab:3}
\end{table}

The Table~\ref{tab:3} shows that for low-lying excited states, the calculations conducted allow for the identification of frequency shifts to transitions between sublevels of fine structure with good accuracy (the numerical calculation error is indicated in parentheses for each value). The numerical calculation error increases not only with the rise of the principal quantum number but also with the increase of the orbital angular momentum. Finally, the signs of the thermal shift for $np$ states and states with $l>1$ are opposite. The continuation of Table~\ref{tab:3} for states with $n=6,7$ reveals that the error in determining the shifts of the fine splitting exceeds the value itself. However, with further increases in the principal quantum number, the calculations return to "normal" behavior. It is worth noting that the sign of the thermal shift for a specific state changes precisely for these states, which, in our opinion, explains the mentioned behavioral traits. In other words, a change in the sign of the level shifts (crossing through zero) is accompanied by a reduction in the thermal shift of the fine structure.

Below we present a series of values for the hydrogen atom and several hydrogen-like ions for the $1s-2s$ and $1s-3s$ transition frequencies as a function of temperature. It is well known that the Stark thermal shift for certain transition frequencies can vanish at the magic temperature \textcolor{black}{:}\cite{PhysRevLett.87.270801,PhysRevA.100.023417}. On the other hand, such a cancellation of the leading order shift should be verified for the possibility of residual contributions (in our case, relativistic effects \cite{Lopez_2025}). The results of calculations for temperatures corresponding to laboratory conditions are summarized in Table~\ref{tab:4}, which also provides the values of $T_m$ at which the thermal frequency shift vanishes.
\begin{widetext}
\begin{center}
\begin{table}[ht]
\centering
\caption{The real part of the thermal one-loop correction to the $1s_{1/2}-2s_{1/2}$ and $1s_{1/2}-3s_{1/2}$ transition frequencies in the hydrogen atom and He$^+$, Li$^{2+}$ ions. The values are given in Hz at different temperatures. The magic temperature, $T_m$, at which the frequency shift becomes zero is given in the last column.
}
\begin{tabular}{l c c c c c c | c} 
\hline\hline
 & & H  & $\Re \Delta \varepsilon_{2s_{1/2}}^{\beta}-\Re \Delta \varepsilon_{1s_{1/2}}^{\beta}$ &  & & & $T_m$\\
\hline
 $40$ K & $50$ K & $60$ K & $70$ K & $80$ K & $90$ K & $100$ K & $65.73(2)$ K\\ 
\hline 
$0.00053(1)$ & $0.00056(1)$  & $0.00034(3)$ & $-0.00032(1)$ & $-0.00162(2)$ & $-0.00372(5)$ & $-0.00694(4)$ & $0.0$\\  

\hline\hline
 & & H  & $\Re \Delta \varepsilon_{3s_{1/2}}^{\beta}-\Re \Delta \varepsilon_{1s_{1/2}}^{\beta}$ &  & & & $T_m$\\

\hline
 $10$ K & $20$ K & $30$ K & $40$ K & $45$ K & $50$ K & $60$ K & $29.83(1)$ K\\ 
\hline
$0.0000847(7)$	& $0.0002118(5)$  &   $-0.000012(2)$ & $-0.001219(1)$ & $-0.002467(2)$ & $-0.004322(4)$ & $-0.01047(1)$ & $0.0$\\
\hline
\hline 

 & & He$^{+}$  & $\Re \Delta \varepsilon_{2s_{1/2}}^{\beta}-\Re \Delta \varepsilon_{1s_{1/2}}^{\beta}$ &  & & & $T_m$\\
 \hline
 $50$ K & $100$ K & $150$ K & $200$ K & $300$ K & $400$ K & $500$ K & $528.9(5)$ K\\ 
\hline
$0.005174$	& $0.02051$  &   $0.044185$	& $ 0.073315$ & $0.130660$ & $0.146473$ & $0.0562175$ & $0.0$\\
\hline\hline 

 & & He$^{+}$  & $\Re \Delta \varepsilon_{3s_{1/2}}^{\beta}-\Re \Delta \varepsilon_{1s_{1/2}}^{\beta}$ &  & & & $T_m$\\
\hline
$20$ K & $50$ K & $100$ K & $150$ K & $200$ K & $300$ K & $400$ K & $238.9(7)$ K\\ 
\hline
$0.0014909$	& $0.00908469$  &   $0.0314011$	& $0.0518514$ & $0.0452887$ & $-0.200031$ & $-1.108767$ & $0.0$\\
\hline\hline

 & & Li$^{2+}$  &   $\Re \Delta \varepsilon_{2s_{1/2}}^{\beta}-\Re \Delta \varepsilon_{1s_{1/2}}^{\beta}$ &  & & & $T_m$\\
\hline
$10$ K	& $20$ K  & $30$ K & $40$ K & $50$ K & $100$ K & $300$ K & $16.44(3)$ K\\   
\hline
$-8.20396\times 10^{-5}$	& $0.000253187$  &   $0.001765346$	& $0.00444289$ & $0.008219243$ & $0.04247243$ & $ 0.413515$ & $0.0$\\
\hline\hline

 & & Li$^{2+}$  &   $\Re \Delta \varepsilon_{3s_{1/2}}^{\beta}-\Re \Delta \varepsilon_{1s_{1/2}}^{\beta}$ &  & & & $T_m$\\
\hline
$10$ K	& $50$ K  & $100$ K & $200$ K & $300$ K & $400$ K & $500$ K & $>600$ K\\   
\hline
$0.00041204$	& $0.02027327$  &   $0.083096$	& $0.320369$ & $0.6632291$ & $1.032194$ & $1.3159493$ & $--$\\
\hline
\hline 
\end{tabular}
\label{tab:4}
\end{table}
\end{center}
\end{widetext}

\begin{figure}[ht!]
\centering
\includegraphics[width=1.0\columnwidth]{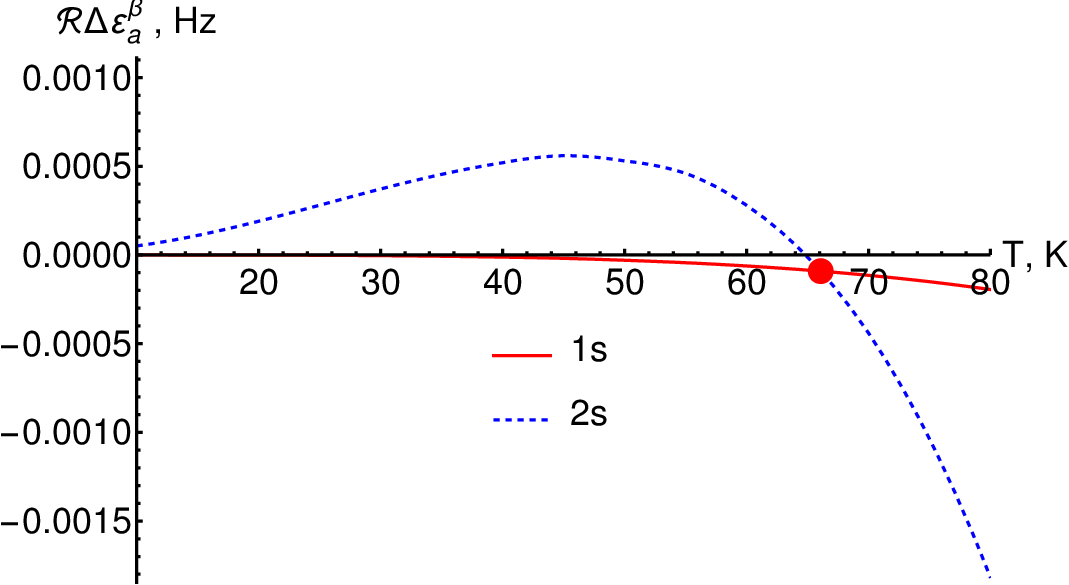}
\caption{The intersection point for the thermal shifts of the states $1s_{1/2}$ (solid line, red online) and $2s_{1/2}$ (dashed line, blue online) in hydrogen, $T_m=65.732(24)$ K. The shift values are plotted in Hz as a function of temperature (in Kelvin). The results were obtained without considering hyperfine structure. 
}
\label{fig:3}
\end{figure}

\begin{figure}[ht!]
\centering
\includegraphics[width=1.0\columnwidth]{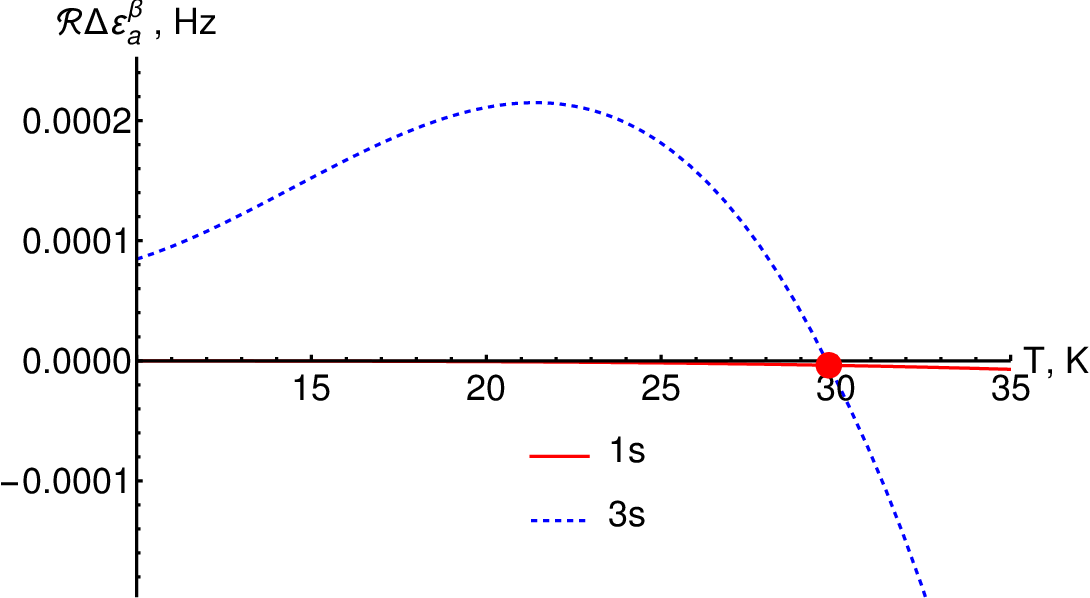}
\caption{The intersection point for the thermal shifts of the states $1s_{1/2}$ (solid line, red online) and $3s_{1/2}$ (dashed line, blue online) in hydrogen, $T_m=29.826(13)$ K. The shift values are plotted in Hz as a function of temperature (in Kelvin). The results were obtained without considering hyperfine structure. 
}
\label{fig:4}
\end{figure}

For greater clarity on the existence of the magic temperature, the following graphs illustrate the thermal shifts for $1s$, $2s$, see Fig~\ref{fig:3}, and for $1s$, $3s$, see Fig.~\ref{fig:4}. Their intersection points show that at this temperature, there will be a zero frequency shift. The corresponding values in the hydrogen atom are $T_m=65.73(2)$ K for $1s-2s$ and $T_m=29.83(1)$ K for $1s-3s$ transition frequencies.

Calculations that take into account the hyperfine structure, even within the framework of Dirac’s theory, require the additional introduction of nuclear spin; see Section~\ref{tr_hfs}. Numerical results of the thermal shifts of the hyperfine transition frequencies in the hydrogen atom and several hydrogen-like ions are presented in Table~\ref{tab:5} at a temperature of $300$ K.
\begin{table}[ht!]
\centering
\caption{Thermal shift to transition frequencies between hyperfine structure levels in a hydrogen atom and some hydrogen-like ions, $\delta\Re \Delta \varepsilon_{\rm HFS}^{\beta} = \delta\Re \Delta \varepsilon_{a}^{\beta}-\delta\Re \Delta \varepsilon_{a'}^{\beta}$. Values are given in Hz at a temperature of $T=300$ K.
}
\resizebox{\columnwidth}{!}{
\begin{tabular}{l c  l c} 
\hline\hline
Transition & $\delta\Re \Delta \varepsilon_{\rm HFS}^{\beta}$ & Transition & $\delta\Re \Delta \varepsilon_{\rm HFS}^{\beta}$ \\ 
\hline
H: $1s_{1/2}^{F=1}-1s_{1/2}^{F=0}$	& $1.3889\times 10^{-8}$   &   $2s_{1/2}^{F=1}-2s_{1/2}^{F=0}$	& $1.73632 \times 10^{-9}$ \\  

H: $2p_{1/2}^{F=1}-2p_{1/2}^{F=0}$	& $6.4289\times 10^{-11}$  &   $2p_{3/2}^{F=2}-2p_{3/2}^{F=1}$	& $8.5658\times 10^{-11}$   \\
\hline
$^3$He$^{+}$: $1s_{1/2}^{F=1}-1s_{1/2}^{F=0}$	& $8.47293 \times 10^{-8}$  &   $2s_{1/2}^{F=1}-2s_{1/2}^{F=0}$	& $1.059379 \times 10^{-8}$ \\  
\hline
$^7$Li$^{2+}$: $1s_{1/2}^{F=2}-1s_{1/2}^{F=1}$	& $2.43206\times 10^{-7}$  &  $^{13}$C$^{5+}$: $1s_{1/2}^{F=1}-1s_{1/2}^{F=0}$	& $ 7.5549\times 10^{-7}$ \\  

\hline\hline

\end{tabular}
}
\label{tab:5}
\end{table}

To perform the numerical calculations, see Eq.~(\ref{se_hfs_final}), the values for the hyperfine splitting were borrowed from Refs.~\cite{HFS_Shabaev_1994,HH-tab,HFS_He+,PhysRevA.108.052802,PhysRevA.52.3686}. First, what follows from Table~\ref{tab:5} is the cubic decrease of the thermal frequency shift with an increase in the principal quantum number (easily verifiable for $s$-states). Secondly, it can be noted that the thermal shift actually increases cubically with the growth of the nuclear charge $Z$, as well as quadratically with the increase in temperature approaching laboratory conditions. 

The corresponding estimate arises from formula (\ref{se_hfs_final}), considering that at room temperature, the thermal radiation frequency $\omega$ is on the order of $2.82\beta\approx 0.00268$ in atomic units, while for hydrogen $\Delta\varepsilon_{\rm HFS}\approx 2.158\times 10^{-7}$ a.u. Then, expanding expression (\ref{se_hfs_final}) in a series with respect to the small parameter $\Delta\varepsilon_{\rm HFS}$, one can obtain $\Delta\varepsilon_{\rm HFS}^\beta\sim \Delta\varepsilon_{\rm HFS}(k_{\rm B}T)^2$, $\Delta\varepsilon_{\rm HFS}\sim Z^3$ \cite{HFS_Shabaev_1994,Grespo-hfs}. However, this estimate is valid only up to $Z=23$. For larger values of the nuclear charge, it is necessary to perform a decomposition by the frequency of photons $\omega$ from thermal radiation. Then, a different dependence will emerge, and moreover, one of the opposite sign: $\Delta\varepsilon_{\rm HFS}^\beta\sim -(k_{\rm B}T)^4 / \Delta\varepsilon_{\rm HFS}$. In the intermediate region, the complete form of equation~(\ref{se_hfs_final}) should be considered. In certain specific cases, the effect of finite lifetime can play a role \cite{Lopez_2025}.

\begin{figure}[t]
\centering
\includegraphics[width=1.0\columnwidth]{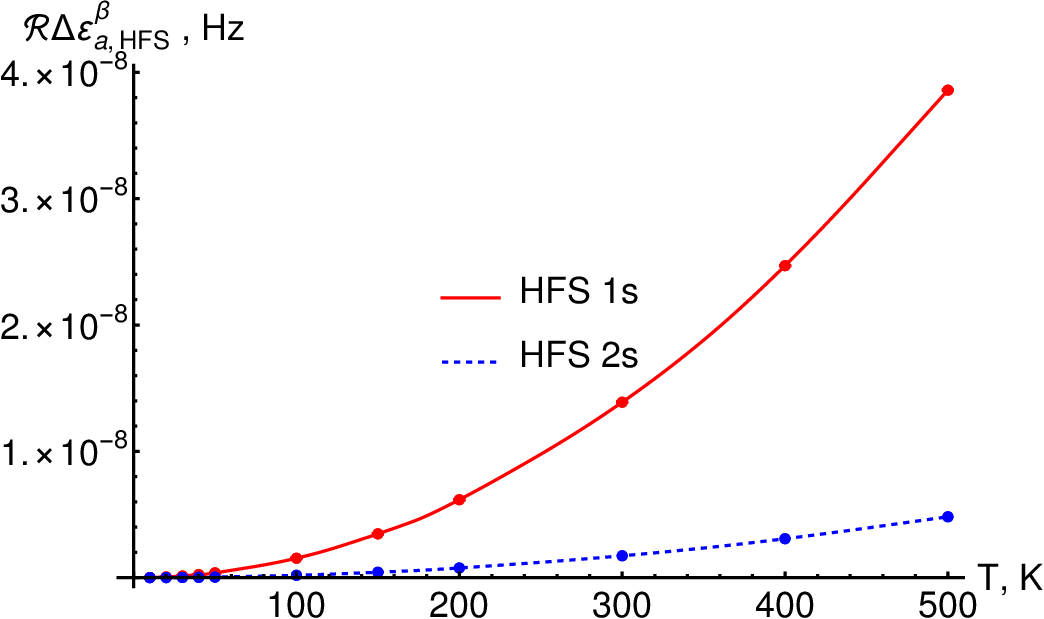}
\caption{Thermal shift of the hyperfine transition frequency, corresponding to the real part of the one-loop self-energy correction for a bound electron, for various temperatures and the $1s_{1/2}$ (solid line, red online), $2s_{1/2}$ (dashed line, blue online) states in the hydrogen atom.}
\label{fig:5}
\end{figure}

The temperature dependence of the thermal shift of the hyperfine $1s$, $2s$ transition frequency for the hydrogen atom is shown in Fig.~\ref{fig:5}. For ions with larger $Z$, the graphs are similar at relevant temperatures and are not presented for brevity.
The corresponding values are collected in Table~\ref{tab:6}.
\begin{widetext}
\begin{center}
\begin{table}[ht!]
\centering
\caption{The real part of the thermal one-loop correction to the frequencies of the lower transitions between the hyperfine structure levels in the hydrogen atom and some H-like ions. The values are given in Hz at a temperature of $T=300$ K.
}
\resizebox{\columnwidth}{!}{
\begin{tabular}{l c c c c c} 
\hline\hline
$\Delta \varepsilon_{a}^\beta$, Hz & $40$ K & $50$ K & $100$ K & $200$ K & $300$ K \\ 
\hline
$1s_{\rm HFS}$ in H  & $2.468659(1)\times 10^{-10}$ & $ 3.857506(2)\times 10^{-10}$ & $1.543182(1)\times 10^{-9}$ & $ 6.17308(1)\times 10^{-9}$ & $1.388970(2)\times 10^{-8}$ \\

$2s_{\rm HFS}$ in H  & $3.0860321(9)\times 10^{-11}$ & $4.822198(1)\times 10^{-11}$ & $1.9290977(8)\times 10^{-10}$ & $7.716828(1)\times 10^{-10}$ & $1.7363190(9)\times 10^{-9}$ \\
\hline
\hline 
$1s_{\rm HFS}$ in $^3$He$^+$  & $1.505650(5)\times 10^{-9}$ & $2.35286(1)\times 10^{-9}$ & $9.41336(3)\times 10^{-9}$ & $3.76565(6)\times 10^{-8}$ & $8.47293(5)\times 10^{-8}$ \\

$2s_{\rm HFS}$ in $^3$He$^+$  & $1.882871(1)\times 10^{-10}$ & $2.942156(1)\times 10^{-10}$ & $1.1769978(9)\times 10^{-9}$ & $4.708260(9)\times 10^{-9}$ & $1.059379(1)\times 10^{-8}$ \\

\hline
\hline 
$1s_{\rm HFS}$ in $^7$Li$^{2+}$  & $4.31508(4)\times 10^{-9}$ & $6.7467(1)\times 10^{-9}$ & $2.70128(2)\times 10^{-8}$ & $1.08084(6)\times 10^{-7}$ & $2.43206(5)\times 10^{-7}$ \\
\hline\hline

$1s_{\rm HFS}$ in $^{13}$C$^{5+}$  & $1.33034(3)\times 10^{-8}$ & $2.0852(1)\times 10^{-8}$ & $8.3799(2)\times 10^{-8}$ & $3.3567(4)\times 10^{-7}$ & $7.5549(4)\times 10^{-7}$ \\
\hline\hline
  
\end{tabular}
}
\label{tab:6}
\end{table}
\end{center}
\end{widetext}

\begin{figure}[ht!]
\centering
\includegraphics[width=1.0\columnwidth]{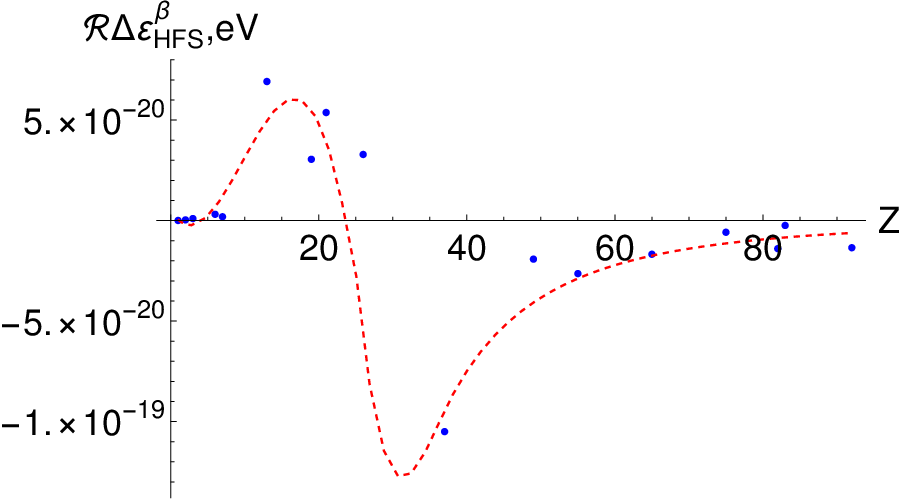}
\caption{Thermal frequency shift of the hyperfine transitions for the ground state of the hydrogen atom and hydrogen-like ions as a function of nuclear charge $Z$. \textcolor{black}{The dashed line represents the interpolation of the calculated values according to the asymptotic behaviors $Z^3$ for $Z<23$ and $1/Z^3$ for larger values of the nuclear charge.} The values of frequency shift are given in eV. 
}
\label{fig:6}
\end{figure}

The value of the thermal shift to the hyperfine transition frequency in the hydrogen atom agrees well with the results of \cite{Itano}, but it is negligibly small even with respect to the accuracy of the experiment of \cite{Hellwig}. However, for neutral and singly ionized atoms, the thermal shift of the hyperfine splitting remains the same in terms of the ratio of the shift to the transition frequency and undoubtedly makes a significant contribution to the budget of experimental error \cite{PhysRevA.110.043108,Lopez_2025}. Below we present a graph in Fig.~\ref{fig:6} illustrating the $Z$-dependence of the shift $\Delta\varepsilon_a^{\beta}$ expressed by Eq.~(\ref{se_hfs_final}).

\begin{figure}[ht!]
\centering
\includegraphics[width=1.0\columnwidth]{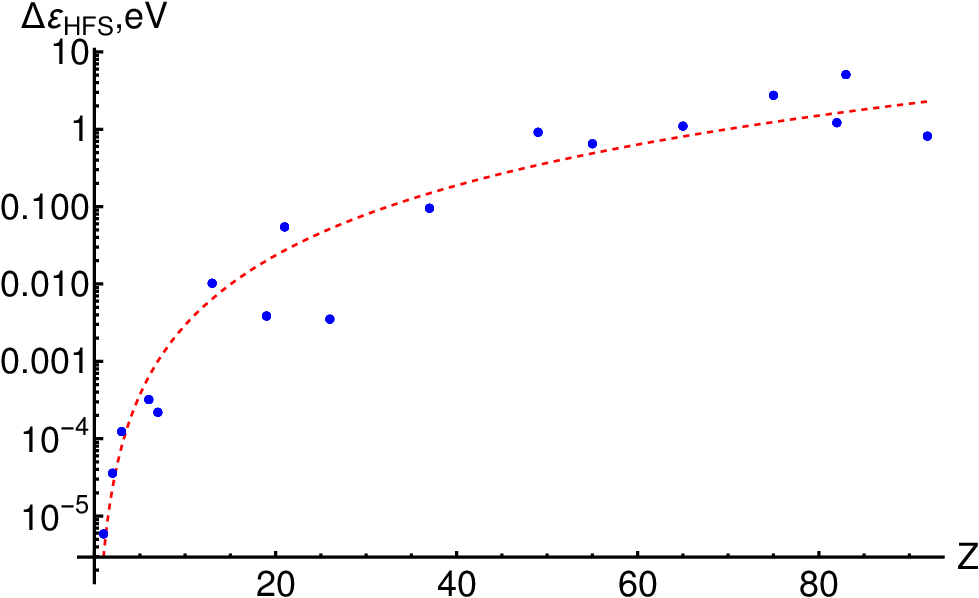}
\caption{The graph shows the dependence of the hyperfine splitting interval of the ground state in hydrogen-like ions on the nuclear charge $Z$ (in logarithmic scale). The points represent the used values, $\Delta\varepsilon_{\rm HFS}$ in eV. The dashed line corresponds to the dependence $\frac{1}{2}\Delta\varepsilon_{\rm HFS}^{Z=1} Z^3$, chosen as the interpolation according to \cite{HFS_Shabaev_1994}, see Table~\ref{tab:7} for details.}
\label{fig:7}
\end{figure}

It should be emphasized that all deviations from asymptotic behavior for $Z<23$ and $Z>23$ are fully correlated with the values of the hyperfine structure interval, see Fig.~\ref{fig:7}. 
Numerical values for $\Delta\varepsilon_{\rm HFS}$ and $\Delta\varepsilon_{\rm HFS}^\beta$, calculated using equation~(\ref{se_hfs_final}) at a temperature of 300 K, are collected in Table~\ref{tab:7}.
\begin{table}[ht!]
\centering
\caption{Thermal shift for hyperfine transition frequencies of the ground state as a function of nuclear charge $Z$. \textcolor{black}{The energy intervals used are indicated in the third column, and the frequency thermal shift is in the fourth column.} All values are given in eV at a temperature of $T=300$ K.}
\begin{tabular}{lp{1cm}lr} 
\hline\hline
Ion & $I$ & $\Delta \varepsilon_{\rm HFS}$, eV & $\Delta \varepsilon_{\rm HFS}^{\beta}$, eV \\ 
\hline
$^{1}$H & $1/2$ & $5.874 \times 10^{-6}$ &      $ 5.744 \times 10^{-23}    $  \\
$^{3}$He$^{1+}$ & $1/2$ & $3.584 \times 10^{-5}$ &     $ 3.504 \times 10^{-22}    $  \\
$^{7}$Li$^{2+}$ & $3/2$ & $1.235 \times 10^{-4}$ &     $ 1.006 \times 10^{-21}    $  \\
$^{13}$C$^{5+}$ & $1/2$ & $3.200 \times 10^{-4}$ &     $ 3.124 \times 10^{-21} $  \\
$^{14}$N$^{6+}$ & $1$ & $2.194 \times 10^{-4}$ &     $ 1.903 \times 10^{-21}    $  \\
$^{27}$Al$^{12+}$ & $5/2$ & $1.022 \times 10^{-2}$ &   $ 6.919 \times 10^{-20}  $  \\
$^{39}$K$^{18+}$ & $3/2$ & $3.853 \times 10^{-3}$ &    $ 3.049 \times 10^{-20} $  \\
$^{45}$Sc$^{20+}$ & $7/2$ & $5.461 \times 10^{-2}$ &   $ 5.376 \times 10^{-20}  $  \\
$^{57}$Fe$^{25+}$ & $1/2$ & $3.511 \times 10^{-3}$ &   $ 3.290 \times 10^{-20} $  \\
$^{85}$Rb$^{36+}$ & $5/2$ & $9.532 \times 10^{-2}$ &   $-1.050 \times 10^{-19}  $  \\
$^{113}$In$^{48+}$ & $9/2$ & $9.148 \times 10^{-1}$ &   
$-1.919 \times 10^{-20}$  \\
$^{133}$Cs$^{54+}$ & $7/2$ & $6.498 \times 10^{-1}$ &  $-2.638 \times 10^{-20}    $  \\
$^{159}$Tb$^{64+}$ & $3/2$ & $1.099$ &   $ -1.675 \times 10^{-20} $  \\
$^{185}$Re$^{74+}$ & $5/2$ & $2.749$ &   $ -5.810 \times 10^{-21} $  \\
$^{207}$Pb$^{81+}$ & $1/2$ & $1.216$ &   $-1.397 \times 10^{-20}    $  \\
$^{209}$Bi$^{82+}$ & $9/2$ & $5.085$ &   $-2.420 \times 10^{-21}    $  \\
$^{235}$U$^{91+}$ & $7/2$ & $8.163 \times 10^{-1}$ &   $-1.350 \times 10^{-20}    $  \\
\hline\hline
\end{tabular}
\label{tab:7}
\end{table}

Combining the results from Tables~\ref{tab:5}, \ref{tab:6}, and Fig.~\ref{fig:6}, one can conclude that the thermal shift for the hyperfine splitting frequency of the ground state (despite the cubic increase with $Z$) is unlikely to be related to corresponding measurements in highly charged ions, see, e.g., \cite{Grespo-hfs}.

The need for accurate calculation of the thermal shift for hyperfine transitions is emphasized by the search for a magic temperature, where the frequencies for the most precise measured transitions (e.g., $1s-2s$, $1s-3s$ in a hydrogen atom) are not subject to the corresponding displacement, see Table~\ref{tab:4}. Performing an analysis similar to those above, see, e.g., Figs.~\ref{fig:3}, \ref{fig:4}, but now taking into account Eq.~(\ref{se_hfs_final}), it was found that the hyperfine structure is insignificant for this purpose. The obtained value for the hyperfine frequency $1s-2s$ is now $T_m=65.86(8)$ K, i.e., the displacement is less than $1$ K. In turn, for $1s-3s$ transition frequency with the account for hyperfine interval shift, we found $T_m=29.66(1)$ K. Such a small influence of the hyperfine structure is quite expected due to its smallness and the formula for the calculated correction, Eq.~(\ref{se_mp}). Given that the main contribution comes from the dipole polarizability and the corresponding thermal Stark shift, the obtained value of the magic temperature should be obtained in the non-relativistic limit, at least to leading order.

\section{Conclusions}
Previously, in the work \cite{Lopez_2025}, it was shown that this QED correction, generalized to the case of finite temperature, includes thermal corrections known from quantum mechanical theory (Stark and Zeeman effects, higher multi-pole polarizabilities). Moreover, a careful transition to the non-relativistic limit, which restores the specified effects, made it possible to point out that in the thermal one-loop correction there are additional contributions related to relativistic effects that were not previously taken into account. In the same work \cite{Lopez_2025}, it was stated that the thermal self-energy correction of a bound electron can be accurately calculated within a fully relativistic approach; see also \cite{Porsev}. Here is the corresponding analysis using partial decomposition by spherical functions for the thermal self-energy operator.

A fully relativistic approach to calculating such a TQED correction was used to compute thermal shifts for various energy levels in different atoms. The applied method allowed for the most accurate calculations to date, significantly increasing the precision of theoretical predictions, see Tables~\ref{table:re_1s_z_300k}, \ref{table:re_exc_h_300k}. All presented values for thermal shifts contain an error from numerical calculations (indicated in parentheses), primarily arising from the basis set used, the box size, and the frequency integration of the final expressions, Eqs.~(\ref{se_mp}), (\ref{se_hfs_final}). The key point is that the calculations were performed without using the static limit, the frequency dependence in the formulas is maintained in the numerical calculation.

As the first result of a fully relativistic approach, we've calculated the thermal shifts of the ground state for hydrogen and hydrogen-like ions with an arbitrary nuclear charge $Z$. Specifically, from Table~\ref{table:re_1s_z_300k}, it follows that the calculation method is most accurate for light ions (with a small $Z$). Another consequence is the behavior of the thermal shift with respect to $Z$. The decrease as $1/Z^4$ has been verified with high accuracy, see Fig.~\ref{fig:re_1s_z_300k}. The dependence $1/Z^4$ was obtained for the non-relativistic limit and dipole approximation \cite{Lopez_2025}. Therefore, a slight deviation for large $Z$ on the graph in Fig.~\ref{fig:re_1s_z_300k} clearly demonstrates the importance of the relativistic approach in highly charged (H-like in particular) ions. 

Based on the analysis above, we have revised the quantum mechanical values of thermal level shifts in a hydrogen atom \cite{farley}, see Table~\ref{table:re_exc_h_300k}. Deviation from generally accepted values (see corresponding sub-strings in Table~\ref{table:re_exc_h_300k}) is explained by several reasons. First and foremost, the need to account for the Lamb shift \cite{PRA_2015,jetp2022} and the fine structure of atomic levels \cite{reex}. The next reason is the approximation for frequency integration, presented in \cite{farley} and still frequently used today for complex atomic systems \cite{Glukhov_2016,PhysRevA.110.043108}. We emphasize that this work presents fully relativistic calculations of the expressions representing the one-loop thermal shift, which is unequivocally consolidated with quantum mechanical theory \cite{Lopez_2025}.

The achieved precision in calculating thermal shifts of atomic energy levels has made it possible to accurately determine the frequency shift for transitions between fine and hyperfine sub-levels at various temperatures and for various hydrogen-like systems, see Tables~\ref{tab:3}, \ref{tab:5} and \ref{tab:6}. The obtained values were used to study the behavior of shifts in hyperfine intervals as a function of both temperature and $Z$, see Figs.~\ref{fig:5}, \ref{fig:6}.

The most important result of the research presented in this work is the discovery of the magic temperature at which the thermal frequency shift is zero \cite{PhysRevA.100.023417}. For the most accurately measured frequencies in the hydrogen atom, as well as for some hydrogen-like ions, the results are presented in Table~\ref{tab:4}, see also Figs.~\ref{fig:3}, \ref{fig:4}. It should be noted that there is a significant difference in the magic temperature for various hydrogen-like systems. For example, for doubly ionized lithium, the thermal shift of the two-photon $1s-3s$ transition frequency vanishes at a temperature significantly exceeding room temperature. At the same time, for lighter systems, magic wavelengths appear achievable in experiments involving two-photon transitions $1s-2s$, $1s-3s$ (except for the helium ion in the latter case). It should also be noted that the analysis, taking into account the hyperfine structure of atomic levels, revealed that the latter is insignificant for determining $T_m$. This can be seen directly from Table~\ref{tab:6}: the thermal shifts to the hyperfine intervals of the ground and $2s$ states were always positive, differing by approximately an order of magnitude and at the same time negligibly small. For a hydrogen atom and the most accurately measured transition frequencies of $1s-2s$ and $1s-3s$ \cite{CODATA-2021}, the values obtained are $T_m=65.9$ K and $T_m=29.7$ K, respectively.

\textcolor{black}{
It is well established that a "magic wavelength" can be used, at which the shifts of the atomic energy levels are equal and thus cancel in the definition of the transition frequency \cite{PhysRevLett.87.270801}. A similar concept can be applied to the thermal shift, i.e., when the levels between which the transition frequency is measured are shifted by the same amount. According to the obtained results, establishing and controlling this $T_m$ (magic temperature) appears to be feasible, see Table~\ref{tab:4}. In our view, stabilizing the temperature of the ambient radiation (from the chamber walls) in combination with employing the magic wavelength technique to cancel the level shift induced by the trapping laser would allow, at least to the lowest order, the influence of thermal radiation to be avoided. A more complex (as it necessitates the corresponding measurements) but realistic alternative appears to be extrapolating the data obtained at different temperatures to the point where the total shift vanishes.
}

\textcolor{black}{
The generalization of the obtained results to multi-electron systems can be achieved in several ways. One approach is based on a quantum mechanical treatment (now commonly accepted) within second-order perturbation theory, which fully corresponds to the non-relativistic limit of the expressions used here; see also the discussion in \cite{Lopez_2025}. In this case, one typically calculates the dipole polarizability of a multi-electron atom and the corresponding dynamic corrections arising from the series expansion of the dynamic polarizability at the low frequency of the thermal radiation. Another possible way is the use of semi-empirical methods, which allow for an efficient description of multi-electron atoms within a single-electron approximation, see, e.g., \cite{PhysRevA.72.062105, PhysRevLett.94.213002}. In this case, the computational method presented in this work can be employed. Finally, the most complex but rigorously consistent method involves a generalization of the one-electron diagrams to the multi-electron case, see, for instance, \cite{lindgren-rel}. We leave the latter two possibilities for future investigations.
}

\section{Acknowledgments}
The work of the author T.Z. was supported by a grant from the Foundation for the Advancement of Theoretical Physics and Mathematics "BASIS"\, No.~23-1-3-31-1. The work of authors A.B. and D.S. was supported by a grant from the Foundation for the Advancement of Theoretical Physics and Mathematics "BASIS"\, No.~25-1-2-18-1.

\bibliographystyle{apsrev4-1}
\bibliography{ref}

\appendix
\onecolumngrid
\section{Reduced matrix element}
\label{ap:rml}
Let us define a function $g_J$ that naturally arises from the partial expansion of the photon propagator~(\ref{PE}):
\begin{eqnarray}
g_{J} (\omega, r_1, r_2) = [J] \omega j_J (\omega r_<) j_J (\omega r_>),
\end{eqnarray}
where $[a] = 2 a + 1$. Then, for the considered matrix elements one can obtain:
\begin{eqnarray}
\label{I}
\langle ab || I (\omega ) || cd \rangle_J = \int_0^\infty dr_{1} dr_{2} \{ (-1)^{J} K_{J}(\kappa_{a},\kappa_{c}) K_J ( \kappa_{b}, \kappa_{d} ) g_J(\omega, r_1, r_2) A_{ac} (r_1) A_{bd} (r_2) 
\\ 
\nonumber
+ \sum_L (-1)^{L+1} [J] g_{L} (\omega,r_1,r_2) D_{JL, ac} (r_1) D_{JL, bd} (r_2)\}.
\label{M_el_f}
\end{eqnarray}
In the above expressions, coefficients that include the radial part of the wave functions are: 
\begin{eqnarray}
A_{ab} (r_1) = G_{a} (r_{1}) G_{b}(r_{1}) + F_{a}(r_{1}) F_{b} (r_{1}), 
\\
\nonumber
D_{JL,ab} (r_1) = G_{a} (r_{1}) F_{b} (r_{1}) H_{L}^J (\kappa_{a}, -\kappa_{b}) - F_{a}(r_{1}) G_{b}(r_{1}) H_{L}^J (-\kappa_{a}, \kappa_{b}), \\
\nonumber
G_{a}(r) = r g_a(r), \quad F_{a}(r) = r f_a(r).
\end{eqnarray} 
Angular coefficients in Eq. (\ref{I}) are:
\begin{eqnarray}
K_{J}(\kappa_{a},\kappa_{b}) = (-1)^{j_b + 1 \slash 2} \sqrt{[j_a][j_b][l_a][l_b]} {\begin{pmatrix}l_a&J&l_b\\0&0&0\end{pmatrix}} {\begin{Bmatrix}j_a&J&j_b\\l_b&\frac{1}{2}&l_a\end{Bmatrix}}, 
\\ 
\nonumber
H_L^J (\kappa_a, \kappa_b) = (-1)^{l_a} \sqrt{6 [j_a][j_b][l_a][l_b]} {\begin{pmatrix}l_a&L&l_b\\0&0&0\end{pmatrix}} {\begin{Bmatrix}j_a&\frac{1}{2}&l_a\\J&1&L\\j_b&\frac{1}{2}&l_b\end{Bmatrix}}.
\end{eqnarray}

\end{document}